\documentclass[11pt,a4paper]{article}

\usepackage[utf8]{inputenc}
\usepackage[T1]{fontenc}
\usepackage[UKenglish]{babel}

\usepackage{amsmath,mathtools,amssymb,amsfonts,bbm}
\usepackage[top=3.11cm,bottom=4.11cm,right=3.11cm,left=3.11cm]{geometry}

\usepackage{physics}
\usepackage{slashed}
\usepackage{tensor}
\usepackage{xcolor,graphicx}
\usepackage{bbold}
\usepackage{appendix}
\usepackage{simplewick}
\usepackage{textpos}
\usepackage{stackrel}
\numberwithin{equation}{section}
\usepackage[style=ieee, citestyle=numeric-comp, backend=biber]{biblatex}
\usepackage{hyperref} % Must usually be the last package to be loaded, but before cleveref.
\usepackage[capitalise]{cleveref}

\usepackage{tabularx}
\newcolumntype{C}{>{\centering\arraybackslash}X}
\newcolumntype{R}{>{\raggedleft\arraybackslash}X}
\newcolumntype{L}{>{\raggedright\arraybackslash}X}
\usepackage{multirow, bigdelim}

\allowdisplaybreaks

\newcommand{\intx}{\int d^4x}
\newcommand{\Dintx}{\int d^Dx}

\newcommand{\hypR}{\mathcal{Y}_R}

\newcommand{\projR}{\mathbb{P}_{\mathrm{R}}}
\newcommand{\projL}{\mathbb{P}_{\mathrm{L}}}

\title{Four-Loop Renormalisation of Chiral Gauge Theories with Non-Anticommuting $\gamma_5$ in the BMHV Scheme}

\newcommand{\email}{}
\author{
Andreas von Manteuffel,$^a$\thanks{\email{manteuffel@ur.de}}\quad
Dominik  Stöckinger,$^b$\thanks{\email{dominik.stoeckinger@tu-dresden.de}}\quad
Matthias Weißwange,$^b$\thanks{\email{matthias.weisswange@tu-dresden.de}}\\[2em]
$^a$Institut für Theoretische Physik, Universität Regensburg,\\ Universitätstraße 31, DE-93040 Regensburg, Germany\\[0.5em]
$^b$Institut für Kern- und Teilchenphysik, TU Dresden,\\ Zellescher Weg 19, DE-01069 Dresden, Germany}

\begin{document}
\thispagestyle{empty}

\maketitle

\setcounter{footnote}{0}
\vspace{2ex}
\begin{abstract}
We present the complete 4-loop renormalisation of an Abelian chiral gauge theory in the Breitenlohner-Maison / `t Hooft-Veltman (BMHV) scheme.
Employing a non-anticommuting $\gamma_5$ in dimensional regularisation, we determine the full set of symmetry restoring counterterms from the quantum action principle.
Our calculation represents the highest-order application of the BMHV framework so far, pushing the limits of a self-consistent treatment of $\gamma_5$ at the multi-loop level.
We describe the computational setup that we developed to perform the computations and discuss key implementation aspects, such as the BMHV algebra and tensor reduction.
Our work demonstrates the feasibility of applying the BMHV scheme at high loop orders and establishes a solid foundation for future studies of high-precision electroweak physics.
\end{abstract}

\newpage
\setcounter{page}{1}

\tableofcontents\newpage

%%%%% Main %%%%%

% +++++++++++++++++++++++++++++++++++++++++++++++++++++++++++++++++++++++++
\section{Introduction}

Chiral gauge theories play a central role in the Standard Model (SM)
and beyond, yet their consistent treatment represents a 
long-standing challenge due to the well-known $\gamma_5$-problem.
In Dimensional Regularisation (DReg), the 
Breitenlohner-Maison/’t Hooft-Veltman (BMHV) scheme, see Refs.\ 
\cite{tHooft:1972tcz,Breitenlohner:1975hg,Breitenlohner:1976te,Breitenlohner:1977hr}
and Refs.\ \cite{Jegerlehner:2000dz,Belusca-Maito:2023wah} for reviews,
provides the only known way for a mathematically consistent treatment 
of $\gamma_5$.
In particular, $\gamma_5$ remains manifestly 4-dimensional
in the BMHV scheme of DReg,
while anticommutativity with the fully $D$-dimensional
$\gamma^{\mu}$-matrices is abandoned.
This results in modified algebraic relations and ultimately in a violation
of gauge and BRST invariance.
Hence, consistency comes at the cost of a regularisation-induced symmetry breaking,
which must be restored in the course of renormalisation.
This leads to an increased computational effort, which becomes particularly
significant in multi-loop applications.

In order to avoid this symmetry breaking, different
$\gamma_5$-schemes have been proposed.
For example, ``Kreimer's scheme'', Refs.\
\cite{Kreimer:1989ke,Korner:1991sx,Kreimer:1993bh}, abandons 
the cyclicity of $\gamma$-traces, while ``Larin's prescription'', Ref.\ \cite{Larin:1993tq}, 
suggests a treatment for axial currents in specific applications.
However, none of these approaches constitutes a fully self-consistent scheme
that is free of ambiguities.
In particular, their scope of application at the 
multi-loop level is limited and not entirely
under control, as demonstrated in e.g.\ Refs.\ \cite{Chetyrkin:2012rz,Zoller:2015tha,Bednyakov:2015ooa,Poole:2019txl,Davies:2019onf,Davies:2021mnc,Herren:2021vdk,Chen:2023lus,Chen:2024zju}.
Therefore, we adopt a no-compromise approach to the treatment of 
$\gamma_5$ within the BMHV scheme.

The BMHV scheme has been studied at the 1-loop
level in non-Abelian chiral gauge theories with 
and without scalar fields in Refs.\ \cite{Belusca-Maito:2020ala,OlgosoRuiz:2024dzq},
and \cite{Martin:1999cc,Cornella:2022hkc}, respectively,
and in the Abelian Higgs model in Ref.\ \cite{Sanchez-Ruiz:2002pcf}.
A general Abelian chiral gauge theory
with scalars as well as different choices for evanescent gauge interactions 
and dimensionally regularised fermions was analysed at the 1-loop level 
in Ref.\ \cite{Ebert:2024xpy}.
Recent applications of the BMHV scheme to effective field theories can be found in Refs.\ \cite{Naterop:2023dek,DiNoi:2023ygk,Heinrich:2024ufp,Heinrich:2024rtg,Naterop:2024ydo}.
In Abelian theories, the BMHV scheme has thoroughly been studied
at higher orders, with a 2-loop application in Refs.\ 
\cite{Belusca-Maito:2021lnk,Belusca-Maito:2022wem}
and an extension to the 
3-loop level in Ref.\ \cite{Stockinger:2023ndm}.
For non-Abelian theories, a full 2-loop renormalisation is presented 
in Ref.\ \cite{Kuhler:2025znv}, with initial discussions in Ref.\ \cite{Kuhler:2024fak}. 

The intricacies of the $\gamma_5$-problem as well as inconsistencies 
of other $\gamma_5$-schemes
that emerge at the multi-loop level render the multi-loop 
properties of the BMHV scheme highly relevant.
In this work, we present the first 4-loop application of the BMHV scheme,
marking the highest-order calculation performed within this framework
and pushing the limits of our approach to symmetry restoration,
see Refs.\ \cite{Belusca-Maito:2020ala,Belusca-Maito:2021lnk,Stockinger:2023ndm,Kuhler:2024fak,Ebert:2024xpy}.
This approach is based on the quantum action principle of DReg, 
see Ref.\ \cite{Breitenlohner:1977hr} and 
Ref.\ \cite{Belusca-Maito:2023wah} for a review. 
In particular, this work directly extends 
previous studies in Refs.\
\cite{Belusca-Maito:2021lnk,Stockinger:2023ndm}.
Performing such multi-loop calculations 
requires an efficient computational framework.
To address these challenges, we developed a completely new computational setup,
where most computationally expensive operations are carried out in 
\texttt{FORM} \cite{Vermaseren:2000nd,Ruijl:2017dtg,FORM:Manual}.
This significantly enhances computational performance, 
enabling not only the current 4-loop calculation,
which involves billions of terms in intermediate steps,
but also future SM applications.

This paper is organised as follows:
In Sec.\ \ref{Sec:TheoryDefinition}, we introduce the 
Abelian chiral gauge theory considered in this work.
In Sec.\ \ref{Sec:ComputationalSetup}, we discuss our computational
setup and the methods used to obtain our results, including
the implementation of the BMHV algebra,
the tadpole decomposition for extracting UV divergences, and the tensor reduction.
In Sec.\ \ref{Sec:4-Loop-Renormalisation}, we present our results for the 4-loop 
renormalisation of the model under consideration, including the explicit symmetry restoration.
Finally, we conclude in Sec.\ \ref{Sec:Conclusion}.

% -------------------------------------------------------------------------

% +++++++++++++++++++++++++++++++++++++++++++++++++++++++++++++++++++++++++
\section{Abelian Chiral Gauge Theory in the BMHV Scheme}\label{Sec:TheoryDefinition}

In this paper, we consider a chiral gauge theory of 
right-handed fermions ${\psi_{R}}_i$ 
interacting with the Abelian gauge boson $B^{\mu}$ 
%of a $U(1)$ gauge group
via the hypercharge matrix $\hypR$. %\footnote{With flavour indices $i, j, \ldots$ labelling the fermions of the theory, e.g.\ leptons and quarks.
%%%%%%%%%%%of different generations (and colour in the case of quarks).
%}
%The renormalisation of this model at 2 and 3 loops has been considered in Refs.\ \cite{Belusca-Maito:2021lnk,Stockinger:2023ndm}.
%Here, we consider its renormalisation at the 4-loop level.
The theory is regularised using DReg and the BMHV scheme for the 
treatment of $\gamma_5$. 
The tree-level action $S_{0}$ of this theory is defined via the 
following $D$-dimensional tree-level Lagrangian
\begin{equation}\label{Eq:TheLagrangian}
    \begin{aligned}
        \mathcal{L} = 
        \mathcal{L}_{\mathrm{fermion}} + \mathcal{L}_{\mathrm{gauge}}
        + \mathcal{L}_{\mathrm{ghost+fix}} + \mathcal{L}_{\mathrm{ext}}.
    \end{aligned}
\end{equation}
For properly regularised loop-diagrams, 
we require $D$-dimensional fermion kinetic terms.
%which impacts the treatment of fermions in $D$-dimensions.
In the present case, this means that we need to introduce 
fictitious, sterile\footnote{
Sterile fields do not participate in any interaction.} 
left-handed partner fields ${\psi^{\text{st}}_{L}}_{\!i}$ for each fermion,
such that we can construct Dirac spinors of the form 
$\psi_i={\psi^{\text{st}}_{L}}_{\!i}+{\psi_{R}}_i$.\footnote{
Here, this is our only option, as the theory contains no physical left-handed fermions.}
In this way the fermion kinetic terms read
\begin{align}\label{Eq:LfermionKin}
    \mathcal{L}_{\mathrm{kin,f}} = 
        \overline{\psi}_j i \slashed{\partial} \psi_j =
                  {\overline{\psi_{R}}}_j i
                  \overline{\slashed{\partial}} {\psi_{R}}_j
                  +
                  {\overline{\psi^{\text{st}}_{L}}}_{\!j} i
                  \overline{\slashed{\partial}} {\psi^{\text{st}}_{L}}_{\!j}
                  +
                  {\overline{\psi_{R}}}_j i
                  \widehat{\slashed{\partial}} {\psi^{\text{st}}_{L}}_{\!j}
                  +
                  {\overline{\psi^{\text{st}}_{L}}}_{\!j} i
                  \widehat{\slashed{\partial}} {\psi_{R}}_j,
\end{align}
where all $D$-dimensional quantities have been split into purely 4-dimensional and remaining $(D-4)$-dimensional parts as e.g.~$\gamma^\mu=\overline{\gamma}^\mu+\widehat{\gamma}^\mu$, such that the modified BMHV commutation relations with $\gamma_5$ are $\{\overline{\gamma}^\mu,\gamma_5\}=[\widehat{\gamma}^\mu,\gamma_5]=0$, see Sec.\ \ref{Sec:BMHV-Algebra} for further details.
Clearly, an inevitable consequence of this modified algebra is the mixing of chiralities in the last two terms of Eq.\ \eqref{Eq:LfermionKin}, i.e.\ in the
evanescent part
of the fermion kinetic term.
Including the gauge boson interaction with the physical fermions, 
the full fermionic Lagrangian is
\begin{equation}\label{Eq:Lfermion}
    \begin{aligned}
        \mathcal{L}_{\mathrm{fermion}} = \overline{\psi}_j i \slashed{\partial} \psi_j 
        - g {\hypR}_{ij} \overline{\psi}_i \projL \overline{\slashed{B}} \projR \psi_j,
    \end{aligned}
\end{equation}
where the hypercharge matrix satisfies the anomaly cancellation condition
\begin{equation}\label{Eq:AnomalyCancellationCondition}
    \begin{aligned}
        \mathrm{Tr}(\hypR^3)\equiv0.
    \end{aligned}
\end{equation}
The mismatch between the purely $4$-dimensional and right-handed 
gauge interaction in Eq.\ \eqref{Eq:Lfermion} and the fully 
$D$-dimensional kinetic term including the resulting mixing 
of left- and right-handed fermions with different gauge quantum numbers
causes the violation of gauge and BRST invariance 
--- the main drawback of the BMHV scheme.
% This results inevitably in a violation of gauge and BRST invariance
% --- the main drawback of the BMHV scheme ---
% since the sterile fields have no gauge interactions, and thus their 
% BRST transformations differ from those of the physical fermions.
For more information regarding this issue, we refer 
the reader to Refs.\ 
\cite{Belusca-Maito:2023wah,Belusca-Maito:2020ala,Belusca-Maito:2021lnk,Stockinger:2023ndm,Kuhler:2024fak} 
and in particular to Ref.\ \cite{Ebert:2024xpy} for a detailed analysis
concerning the treatment of fermions in $D$ dimensions in the context of the BMHV scheme.
For the gauge boson, we have
\begin{equation}
    \begin{aligned}
        \mathcal{L}_{\mathrm{gauge}} = - \frac{1}{4} F^{\mu\nu}F_{\mu\nu},
    \end{aligned}
\end{equation}
with field strength tensor $F_{\mu\nu} = \partial_{\mu}B_{\nu} -
\partial_{\nu}B_{\mu}$.
The familiar $R_\xi$-gauge fixing and the Abelian ghosts enter
the $D$-dimensional Lagrangian via
\begin{equation}\label{Eq:Ddim-LGaugeFixing+Ghost}
    \begin{aligned}
        \mathcal{L}_{\mathrm{ghost+fix}} = 
        - \frac{1}{2\xi} (\partial^{\mu}B_{\mu})^2
        - \overline{c} \Box c.
    \end{aligned}
\end{equation}
Finally, in order to work with Slavnov-Taylor identities, 
along with the quantum action principle of DReg, 
we are required to couple the BRST transformations of
the fields 
\begin{equation}\label{Eq:BRST-Trafos}
    \begin{aligned}
        s_D B_{\mu}(x) &= \partial_{\mu} c(x), \qquad
        %s_D\psi_{i}(x) = 
        s_D{\psi_R}_i(x) = - i g c(x) {\hypR}_{,ij} {\psi_R}_j(x), \qquad
        % s_D\overline{\psi}_i(x) &= s_D\overline{\psi_R}_i(x) = - i g \overline{\psi_R}_j(x) c(x) {\hypR}_{,ji} = i g c(x) \overline{\psi_R}_j(x) {\hypR}_{,ji},\\
        s_D c(x) = 0,
        % s_D\bar{c}(x) &= \mathcal{B}(x) = - \frac{1}{\xi} \partial^{\mu}B_{\mu}(x),\\
        % s_D \mathcal{B}(x) &= 0,
    \end{aligned}
\end{equation}
to external sources $\{\rho^{\mu},R^i,\zeta\}$
and take them into account via
\begin{equation}\label{Eq:Ddim-Lext}
    \begin{aligned}
        \mathcal{L}_{\mathrm{ext}} = 
        \rho^{\mu} s_D B_{\mu} 
        + {\overline{R}}{}^{i} s_D {\psi_R}_i + \big(s_D \overline{\psi_R}_i\big) R^{i}
        + \zeta s_D c,
    \end{aligned}
\end{equation}
where $s_D$ denotes the $D$-dimensional BRST operator.
The aforementioned regularisation induced symmetry breaking is then explicitly given by
\begin{equation} \label{Eq:Delta_Hat-Tree_Level_Breaking}
    \begin{aligned}
        \widehat{\Delta} 
        \equiv
        s_D S_{0}
        = - g \,  {\hypR}_{,ij} \Dintx \, c \, 
                    \Big\{
                    \overline{\psi}_i \Big(
                    \overset{\leftarrow}{\widehat{\slashed{\partial}}} \projR 
                    + \overset{\rightarrow}{\widehat{\slashed{\partial}}} \projL
                    \Big) \psi_j 
                    \Big\}
                = \Dintx \, \widehat{\Delta}(x).
    \end{aligned}
\end{equation}

The goal of the present paper is to determine the 4-loop renormalisation of the model. This requires the computation of power-counting divergent Green functions at the 4-loop level. The set of Green functions consists of standard Green functions with physical external fields and Green functions with insertions of the composite operator $\Delta$ and external Faddeev-Popov ghost $c$.

% -------------------------------------------------------------------------

% +++++++++++++++++++++++++++++++++++++++++++++++++++++++++++++++++++++++++
\section{Computational Setup and Methods}\label{Sec:ComputationalSetup}

In this section, we present the computational setup used for 
all calculations in this project and discuss the methods employed.
These methods and our implementation are applicable to various models; 
thus, we keep the discussion largely model-independent, while 
illustrating certain aspects and challenges, where appropriate, with the help of 
the specific theory introduced in Sec.\ \ref{Sec:TheoryDefinition}.

We begin with a concise overview of our workflow, including
the software tools utilised, and highlight 
major challenges specific to calculations involving non-anticommuting $\gamma_5$ 
in the BMHV scheme of DReg. 
Next, we discuss the so-called BMHV algebra and provide implementation 
details in \texttt{FORM} within our setup.
We then outline our approach for extracting UV divergences using a tadpole 
decomposition and describe the implementation of counterterms.
Finally, we explain our tensor reduction procedure along with its implementation.

\subsection{Computational Setup}\label{Sec:Setup}

To calculate Green functions, we generate Feynman diagrams with 
\texttt{QGRAF} \cite{Nogueira:1991ex} and process its output with a 
sequence of \texttt{FORM} programs.
We start this sequence by inserting the Feynman rules into the diagram to obtain an expression for the multi-leg Green function.
Since we wish to extract its UV divergences, we apply a tadpole decomposition to map all Feynman integrals to fully massive, single scale vacuum bubbles.
Details for this step are given below in Sec.\ \ref{Sec:TadpoleDecomposition}.

Our goal is to express the resulting vacuum bubbles in terms of scalar integrals according to
\begin{equation}\label{Eq:IntegralFamily}
    \begin{aligned}
        I(\nu_1,\ldots,\nu_N) = 
        \big(\mu^{4-D}\big)^L 
        \int \Bigg( \prod_{i=1}^{L} \frac{d^Dk_i}{(2\pi)^D} \Bigg)  
        \frac{1}{\mathcal{D}_1^{\nu_1} \cdots \mathcal{D}_N^{\nu_N}},
    \end{aligned}
\end{equation}
where $L$ denotes the loop-order, $D=4-2\epsilon$ 
is the spacetime dimension, and $N=N(L)=L(L+1)/2$ is the number of 
different propagators for vacuum bubbles without external momenta.
Here, the propagators are of the form
\begin{equation}
    \begin{aligned}
        \mathcal{D}_i = q_i^2 - M^2,
    \end{aligned}
\end{equation}
where $q_i$ is a linear combination of loop momenta.
For each loop order considered in this paper, a single choice %for the set 
of propagator momentum configurations
\begin{equation}
    \begin{aligned}
        B_L = \Big\{ q_i \, \big| \, i \in \{1,\cdots,N(L)\}; \, \{q_i^2\} \,\, \text{linearly independent} \Big\}
    \end{aligned}
\end{equation}
is sufficient to cover all Feynman diagrams.
Here, we adopt the convention of Ref.\ \cite{Luthe:2015ngq}, with propagator momenta
shown in Tab.\ \ref{Tab:PropagatorMomentumConfigurations}.
\begin{table}[t]
    \centering
    \begin{tabular}{|c|c|c|c|c|}
        \hline
                 &$B_1$&$B_2$    &$B_3$    &$B_4$        \\ \hline
         $q_1$   &$k_1$&$k_1$    &$k_1$    &$k_1$        \\
         $q_2$   &     &$k_2$    &$k_2$    &$k_2$        \\
         $q_3$   &     &$k_1-k_2$&$k_3$    &$k_3$        \\
         $q_4$   &     &         &$k_1-k_2$&$k_4$        \\
         $q_5$   &     &         &$k_1-k_3$&$k_1-k_4$    \\
         $q_6$   &     &         &$k_2-k_3$&$k_2-k_4$    \\
         $q_7$   &     &         &         &$k_3-k_4$    \\
         $q_8$   &     &         &         &$k_1-k_2$    \\
         $q_9$   &     &         &         &$k_1-k_3$    \\
         $q_{10}$&     &         &         &$k_1-k_2-k_3$\\ \hline
    \end{tabular}
    \caption{Propagator momenta $q_i$ for fully massive single scale vacuum bubbles up to $L=4$ loops, see Ref.\ \cite{Luthe:2015ngq}.}
    \label{Tab:PropagatorMomentumConfigurations}
\end{table}

Due to ambiguities in the assignment of loop momenta, the momenta encountered in the denominators of our diagrams do not automatically match those of the chosen integral families.
To achieve this, we employ the program \texttt{Feynson} \cite{Maheria:2022dsq}, which systematically shifts loop momenta such that only the momentum configurations listed in Tab.\ \ref{Tab:PropagatorMomentumConfigurations} appear in the denominators of the Feynman integrals.

At this stage of the calculation, we usually encounter tensor integrals
that are not of the form \eqref{Eq:IntegralFamily} 
due to loop momenta with open Lorentz indices in the numerator.
To address this, we apply a tensor reduction which expresses tensor integrals in terms of scalar integrals $I(\nu_1,\ldots,\nu_N)$ defined by Eq.\ \eqref{Eq:IntegralFamily}.
Since tensor reduction is among the most computationally expensive 
steps of the calculation, we have implemented an efficient tensor reduction 
procedure in \texttt{FORM}, which is explained in Sec.\ \ref{Sec:TensorReduction}.

Following the tensor reduction, we perform the Dirac algebra.
Specifically, we employ a non-anticommuting $\gamma_5$ in the framework of the BMHV scheme.
For this purpose, we have implemented a dedicated \texttt{FORM} procedure 
that efficiently 
applies the BMHV algebra while leveraging as many built-in \texttt{FORM}
operations as possible.
A corresponding discussion of this procedure is to be found in Sec.\ 
\ref{Sec:BMHV-Algebra}.

As a final step of the calculation, the general scalar integrals 
$I(\nu_1,\ldots,\nu_N)$ are reduced to a finite set of 
master integrals using integration-by-parts (IBP) identities.
For this purpose, we employ the \texttt{C++} version of 
\texttt{FIRE} \cite{Smirnov:2019qkx,Smirnov:2023yhb}.
To ensure that \texttt{FIRE} reduces all integrals
to a minimal set of preferred master integrals, we additionally use
\texttt{Reduze2} \cite{vonManteuffel:2012np} to identify and generate sector symmetry
relations, which is particularly crucial at the 4-loop level.
We also employed \texttt{Kira}
\cite{Maierhofer:2017gsa,Klappert:2020nbg,Klappert:2019emp,Klappert:2020aqs} and the private code \texttt{Finred} based on finite field arithmetic \cite{vonManteuffel:2014ixa,vonManteuffel:2016xki,Peraro:2016wsq} for cross checks and basis changes.
The solutions for the master integrals at the 2- and 3-loop level
have been taken from Refs.\ \cite{Schroder:2005va,Martin:2016bgz},
while the 4-loop master integrals have been obtained from Ref.\
\cite{Czakon:2004bu}.

In our new computational setup, we employ \texttt{Mathematica} \cite{mathematica} only to interface
various software tools and for the automated generation of
\texttt{FORM} code.
All performance-relevant operations are carried out in 
carefully optimised \texttt{FORM} programs and dedicated IBP solvers,
removing a major bottleneck encountered in past calculations, cf.\ Refs.\ \cite{Stockinger:2023ndm,Kuhler:2024fak}.

The setup described above has successfully been tested and
applied up to the 4-loop level.
We performed a complete renormalisation of ordinary 
vector-like quantum electrodynamics (QED)
at the 4-loop level and validated our results against the literature.
Additionally, we successfully reproduced our previous results 
for the Abelian chiral gauge theory considered here up to the 3-loop level, 
see Refs.\ \cite{Belusca-Maito:2021lnk,Stockinger:2023ndm}.

\subsection{BMHV-Specific Challenges}\label{Sec:BMHVSpecificChallenges}

In this subsection, we highlight two major challenges
specific to calculations in chiral gauge theories within the BMHV scheme,
which increase the computational effort compared to vector-like gauge theories.
These challenges arise due to the regularisation-induced symmetry breaking and 
the modified algebraic relations responsible for this breaking,
details for which will be given in Sec.\ \ref{Sec:BMHV-Algebra} below.
To illustrate these challenges, we briefly contrast 
ordinary vector-like QED (regularised using naive DReg)
with the Abelian chiral gauge theory
studied in this work (regularised in the BMHV scheme of DReg),
using explicit examples of Green functions and Feynman diagrams.
Afterwards, we comment on the implications for the computation of physical observables 
and emphasise that, despite these challenges, the BMHV scheme remains of practical relevance.
The two primary computational challenges in the BMHV scheme are:
\begin{enumerate}
    \item[$(i)$] Ward and Slavnov-Taylor identities are broken, preventing their usage.
    \item[$(ii)$] More Lorentz covariants occur, involving both $D$-dimensional and $(D-4)$-dimensional indices.
\end{enumerate}

The first issue implies that Ward or Slavnov-Taylor identities
cannot be used to circumvent the computation of multi-leg ($\geq3$) 
1PI Green functions, as is typically done (see e.g.\ \cite{Zoller:2015tha,Zoller:2016sgq,Chetyrkin:2017mwp,Herzog:2017ohr,Luthe:2017ttg,Davies:2019onf,Davies:2021mnc,Herren:2021vdk}).
Consequently, for a complete renormalisation of a given chiral gauge theory
in the BMHV scheme at a given loop order, 
1PI Green functions with up to five external legs must be calculated, 
see Refs.\ \cite{Ebert:2024xpy,Kuhler:2025znv}.
In the Abelian chiral gauge theory considered here, we need to take 
Green functions with up to four external legs into account,
see Sec.\ \ref{Sec:4-Loop-Ssct} and \ref{Sec:4-Loop-BRST-Breaking}.
%and specifically Eqs.\ \eqref{Eq:GF-BBBB} and \eqref{Eq:GF-cBBB}.
Clearly, this significantly increases the total number of 1PI Green functions
required for a complete renormalisation.

The second issue concerns the proliferation of Lorentz covariants.
On the one hand, it is necessary to compute multi-leg ($\geq3$) 1PI Green functions due to symmetry breaking,
cf.\ $(i)$, leading to more complicated Lorentz structures.
This issue is evident not only in previous results, Refs.\ 
\cite{Belusca-Maito:2020ala,Belusca-Maito:2021lnk,Belusca-Maito:2023wah,Stockinger:2023ndm,Kuhler:2024fak,Ebert:2024xpy},
but also in the present results in 
Sec.~\ref{Sec:4-Loop-Renormalisation}.
On the other hand, the BMHV algebra leads to both $D$-dimensional and $(D-4)$-dimensional Lorentz indices.
%(see Eq.\ \eqref{Eq:BMHV-split})
Our goal is to work with standard, scalar Feynman integrals.
In ordinary vector-like QED one could simply employ projectors at the level of Green functions and contract Lorentz indices to achieve this.
Here, instead, we wish to avoid the contraction of loop momenta with non-$D$-dimensional indices resulting from the evaluation of the BMHV Dirac algebra,
as this can generate $4$- and $(D-4)$-dimensional loop momentum numerator polynomials.
We solve this problem by performing a $D$-dimensional tensor reduction of the tensor integrals.
Since we encounter high tensor ranks, our tensor reduction procedure needs to be highly efficient.

As an alternative to tensor reduction as used in this work, one could employ a method to deal with $(D-4)$ dimensional scalar products  of the form $\widehat{k}_i\cdot\widehat{k}_j$ in the numerators of the loop integrals, also known as ``$\mu$-terms''.
These non-standard loop integrals can not directly be processed by standard IBP solvers and their solutions are not readily available.
They thus require a dedicated treatment.
At 1- and 2-loop order, a suitable method has been discussed in Refs.\ \cite{Bern:1995db,Bern:2002tk} already some time ago.
A review and an application at 2-loops was presented in Ref.\ \cite{Heller:2020owb}.
At higher loop orders, no such method has been worked out yet.
In this paper, we do not consider this option any further and strictly insist on purely $D$-dimensional loop integrals.

To summarise, the renormalisation of chiral gauge theories requires 
a significantly higher computational effort compared to vector-like gauge theories.
This is not only due to the increased number of required Green functions,
but also because each Green function involves higher computational complexity.
Moreover, the regularisation-induced violation of gauge and BRST invariance
necessitates the determination of finite, symmetry-restoring counterterms.
These can be derived from 1PI Green functions with an insertion 
of a local composite operator $\Delta$ that encodes the symmetry breaking, see Sec.\ \ref{Sec:4-Loop-BRST-Breaking} for a brief illustration of our methodology.
Although this symmetry-restoration procedure is highly efficient,
%see Refs.\ \cite{Belusca-Maito:2020ala,Belusca-Maito:2021lnk,Belusca-Maito:2023wah,Stockinger:2023ndm,Kuhler:2024fak,Ebert:2024xpy}, 
it still requires the computation of additional Green functions.

\begin{figure}[t!]
    \centering
    \includegraphics[width=0.38\textwidth]{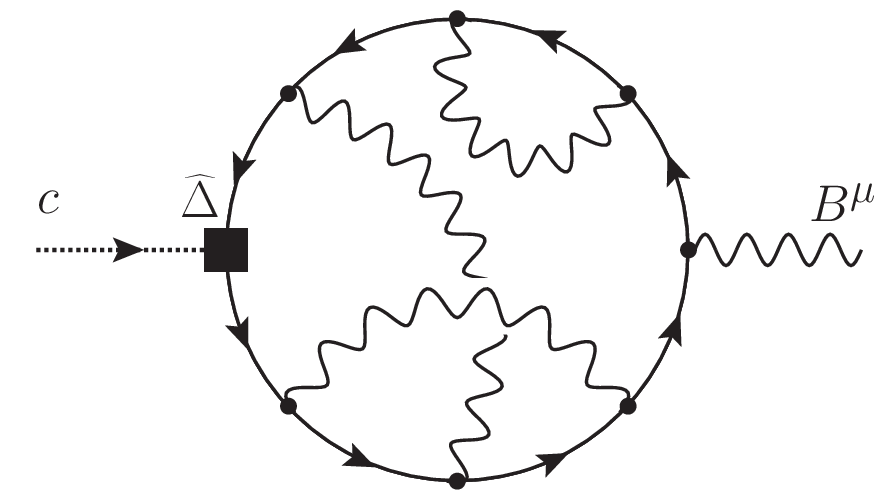}
    \hspace{3.5em}
    \includegraphics[width=0.38\textwidth]{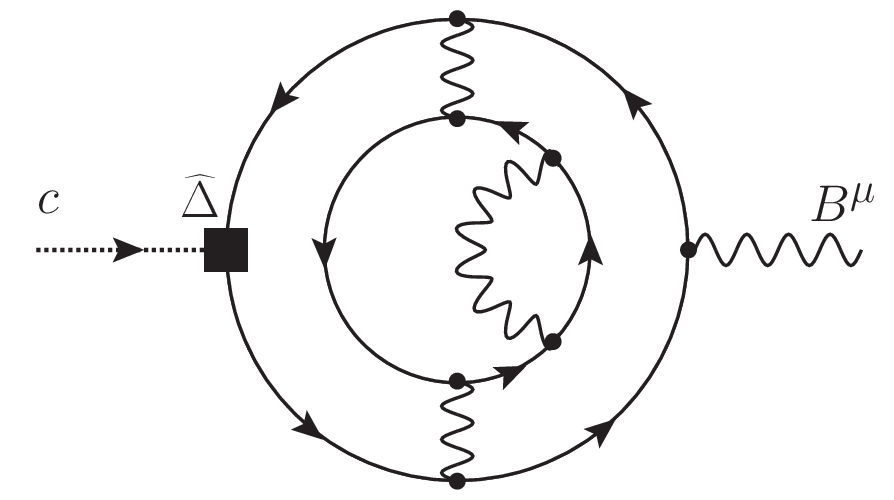}
    \caption{Representative 4-loop Feynman diagrams contributing to 
    the $\Delta$-operator-inserted 1PI Green function $i \Delta \cdot \Gamma|_{B_{\mu}c}^{4}$.}
    \label{Fig:cA-4-Loop-Diagrams}
\end{figure}

\begin{figure}[t!]
    \centering
    \includegraphics[width=0.38\textwidth]{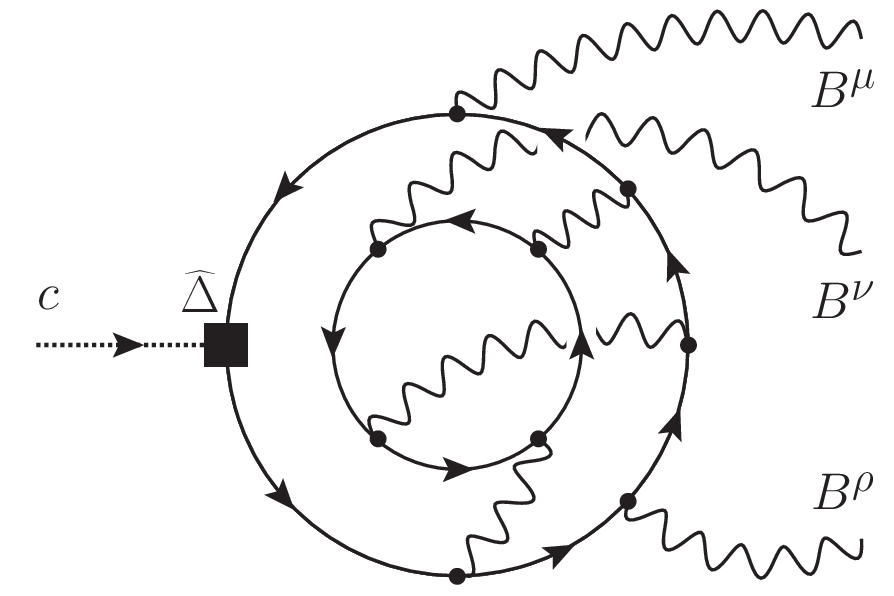}
    \hspace{3.5em}
    \includegraphics[width=0.44\textwidth]{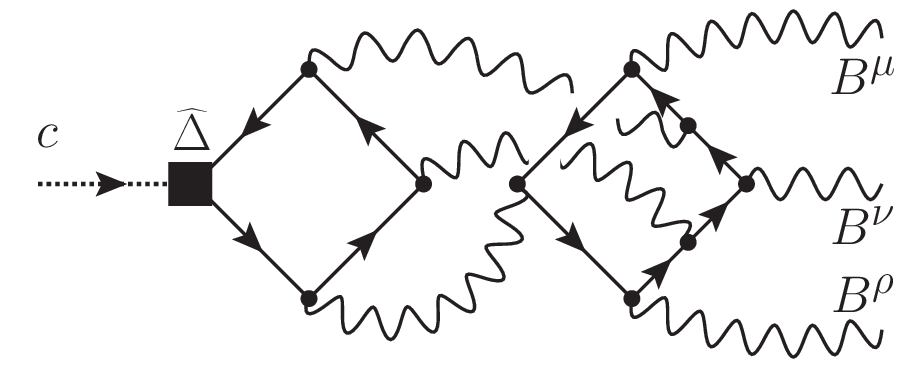}
    \caption{Representative 4-loop Feynman diagrams contributing to 
    the $\Delta$-operator-inserted 1PI Green function $i \Delta \cdot \Gamma|_{B_{\rho}B_{\nu}B_{\mu}c}^{4}$.}
    \label{Fig:cAAA-4-Loop-Diagrams}
\end{figure}

In the following we illustrate these challenges by comparing
the renormalisation of ordinary 
vector-like QED and that of the Abelian chiral gauge
theory studied here, see Sec.\ \ref{Sec:TheoryDefinition}.
In vector-like QED, Ward identities ensure that it is sufficient to compute 
only the self-energies of the gauge boson and the fermions 
in order to obtain all renormalisation constants.
In contrast, the Abelian chiral gauge theory, regularised in the BMHV scheme,
requires the computation of five standard Green functions
and four Green functions with the insertion of a local composite operator $\Delta$.
The required Green functions will be listed below in 
Eqs.\ \eqref{Eq:GF-BB}-\eqref{Eq:GF-BBBB} as well as Eqs.\ \eqref{Eq:GF-cB}-\eqref{Eq:GF-cBBB},
and involve e.g.\ four-point functions $B^{\mu}B^{\nu}B^{\rho}B^{\sigma}$ and 
$cB^{\mu}B^{\nu}B^{\rho}$ as well as an unusual 2-point function $cB^{\mu}$, where
$c$ is the Faddeev-Popov ghost.
Among these, the $\Delta$-inserted Green functions associated to 
$cB^{\mu}$ and $cB^{\mu}B^{\nu}B^{\rho}$ are particularly computationally expensive.
Both involve rank-$12$ tensor integrals at the 4-loop level.
The $cB^{\mu}$ Green function additionally has a superficial degree of divergence of $3$, 
which leads to a large number of terms after tadpole decomposition,
see Sec.\ \ref{Sec:TadpoleDecomposition}.
Two representative 4-loop diagrams for this Green function
are shown in Fig.\ \ref{Fig:cA-4-Loop-Diagrams}.
The $4$-point Green function $cB^{\mu}B^{\nu}B^{\rho}$, while having a comparatively 
low degree of divergence of $1$, involves a significantly larger number Feynman diagrams,
see Fig.\ \ref{Fig:cAAA-4-Loop-Diagrams} for two 4-loop representatives.
All of these diagrams involve the insertion of
$\widehat{\Delta}$, see Eq.~\eqref{Eq:Delta_Hat-Tree_Level_Breaking}, 
which corresponds to the tree-level contribution of 
the composite operator $\Delta$.

\begin{figure}[t!]
    \centering
    \includegraphics[width=0.38\textwidth]{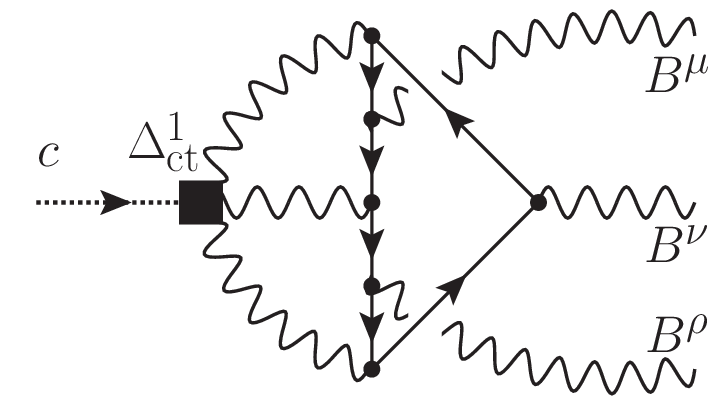}
    \hspace{3.5em}
    \includegraphics[width=0.38\textwidth]{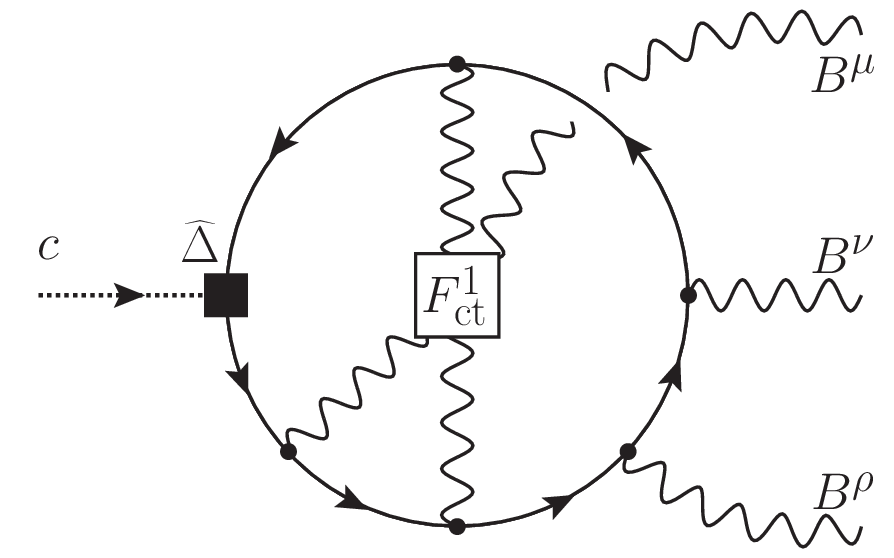}
    \caption{Representative 3-loop Feynman diagrams with counterterm insertions contributing to 
    the 4-loop, subrenormalised, $\Delta$-operator-inserted 1PI Green function $i \Delta \cdot \Gamma|_{B_{\rho}B_{\nu}B_{\mu}c}^{4}$. The left diagram features the insertion of a new $\Delta$-vertex resulting from a 1-loop correction to the $\Delta$-operator, while the right diagram involves the insertion of a 1-loop, finite, symmetry-restoring counterterm.}
    \label{Fig:cAAA-3LCT1-Diagrams}
\end{figure}

The BMHV scheme does not introduce additional tree-level 
vertices which could enter standard Green functions (without $\Delta$-insertions).
Hence, at the $L$-loop level, the number of genuine $L$-loop diagrams
in standard Green functions is the same as in other regularisation schemes.
For example, the 1PI Green functions associated to $B^{\mu}B^{\nu}$, $\overline{\psi}\psi$,
$\overline{\psi}B^{\mu}\psi$ and $B^{\mu}B^{\nu}B^{\rho}B^{\sigma}$ 
contain the same number of genuine $L$-loop diagrams in both vector-like QED 
and the BMHV-regularised Abelian chiral gauge theory.
However, symmetry-restoring counterterms lead to
additional $(<\!L)$-loop counterterm-inserted diagrams.
Moreover, since the breaking $\Delta$ itself receives loop corrections, 
further $(<\!L)$-loop counterterm-inserted diagrams arise from these higher-order contributions.
Fig.\ \ref{Fig:cAAA-3LCT1-Diagrams} shows two representative 3-loop counterterm diagrams 
that contribute to the 4-loop result for the Green function $cB^{\mu}B^{\nu}B^{\rho}$.
The left diagram includes the insertion of a 1-loop correction to the $\Delta$-operator,
while the right diagram features the insertion of a 1-loop symmetry-restoring counterterm.
Both of these counterterm vertices have no tree-level analogue,
highlighting the increased structural complexity introduced by the required 
symmetry-restoration in the BMHV scheme.


Despite these challenges, 
the BMHV scheme remains viable for practical applications.
This is primarily because the full renormalisation procedure
needs to be performed only once.
Once completed, physical observables can be computed just as in any other regularisation scheme,
provided the BMHV algebra is consistently applied.
Importantly, the number of Green functions required for the calculation 
of a given observable does not increase,
nor does the number of diagrams for the genuine $L$-loop contributions.
Furthermore, $\Delta$-vertices do no longer play a role in calculations of observables after the 
symmetry is restored.
Only the number of counterterms is increased,
yet even these remain manageable:
as shown in Sec.\ \ref{Sec:4-Loop-Renormalisation},
the resulting counterterms can be expressed in a compact form, even at the 4-loop level,
making them well-suited for efficient computer implementations.
Hence, the BMHV scheme is not only theoretically consistent, 
but also practically applicable for higher-order computations in chiral gauge theories.

\subsection{BMHV Algebra with Non-Anticommuting $\gamma_5$}\label{Sec:BMHV-Algebra}

In the BMHV scheme, full anticommutativity of $\gamma_5$ is abandoned in $D$ dimensions, 
which results in modified algebraic relations for the $\gamma$-matrices 
--- the source of the regularisation-induced breaking of gauge and BRST invariance, 
see Eq.\ \eqref{Eq:Delta_Hat-Tree_Level_Breaking}.
In particular, the (quasi) $D$-dimensional space of DReg is decomposed into a $4$- and 
a $(D-4)-$dimensional component as $\mathbb{M}=\mathbb{M}_{4}\oplus\mathbb{M}_{D-4}$,
see Refs.\ \cite{Belusca-Maito:2020ala,Belusca-Maito:2021lnk,Stockinger:2023ndm} and Ref.\
\cite{Belusca-Maito:2023wah} for a review,
where $4$- and $(D-4)-$dimensional objects are denoted with 
overbars and hats, respectively.
Thus, all Lorentz covariants admit a decomposition of the form
\begin{equation}\label{Eq:BMHV-split}
    \begin{aligned}
        \gamma^{\mu} = \overline{\gamma}^{\mu} + \widehat{\gamma}^{\mu},
        \qquad
        \eta_{\mu\nu} = \overline{\eta}_{\mu\nu} + \widehat{\eta}_{\mu\nu},
        \qquad
        X^{\mu} = \overline{X}^{\mu} + \widehat{X}^{\mu},
    \end{aligned}
\end{equation}
for any Lorentz-vector $X^{\mu}$ and with
\begin{equation}\label{Eq:BMHV-Metric-Contractions}
    \begin{aligned}
        \overline{\eta}^{\mu\nu}\overline{\eta}_{\mu\nu}=4,
        \qquad
        \widehat{\eta}^{\mu\nu}\widehat{\eta}_{\mu\nu}=D-4,
        \qquad
        \overline{\eta}^{\mu\nu}\widehat{\eta}_{\mu\nu}=0.
    \end{aligned}
\end{equation}
This leads to the aforementioned modification of the $\gamma$-algebra
\begin{equation}\label{Eq:BMHV-Algebra-1}
    \begin{aligned}
        \{\overline{\gamma}^{\mu},\overline{\gamma}^{\nu}\} &= 2 \overline{\eta}^{\mu\nu} \mathbb{1},
        &
        \qquad
        \{\widehat{\gamma}^{\mu},\widehat{\gamma}^{\nu}\} &= 2 \widehat{\eta}^{\mu\nu} \mathbb{1},
        &
        \qquad
        \{\overline{\gamma}^{\mu},\widehat{\gamma}^{\nu}\} &= 0,\\
        \overline{\gamma}_{\mu} \overline{\gamma}^{\mu} &= 4 \mathbb{1},
        &
        \widehat{\gamma}_{\mu} \widehat{\gamma}^{\mu} &= (D-4) \mathbb{1},
        &
        \overline{\gamma}_{\mu} \widehat{\gamma}^{\mu} &= 0,
    \end{aligned}
\end{equation}
where $\gamma_5$, as a manifestly $4$-dimensional quantity,
anticommutes only with the $4$-dimensional components of the 
$\gamma$-matrices, $\overline{\gamma}^{\mu}$,
and commutes with the evanescent components $\widehat{\gamma}^{\mu}$,
such that
\begin{equation}\label{Eq:BMHV-Algebra-2}
    \begin{aligned}
        \{\gamma_5,\overline{\gamma}^{\mu}\} &= 0,
        &
        \qquad
        [\gamma_5,\widehat{\gamma}^{\mu}] &= 0,\\
        [\gamma_5,\overline{\gamma}^{\mu}] &= 2\gamma_5\overline{\gamma}^{\mu},
        &
        \{\gamma_5,\widehat{\gamma}^{\mu}\} &= 2\gamma_5\widehat{\gamma}^{\mu}.
    \end{aligned}
\end{equation}
Further, the $\gamma$-traces are defined such that
\begin{equation}\label{Eq:GammaTraceDefinitions}
    \begin{aligned}
        \mathrm{Tr}(\mathbb{1}) = 4,
        \qquad\qquad
        \mathrm{Tr}(\gamma^{\mu}) = 0.
    \end{aligned}
\end{equation}
In contrast to some other schemes, cyclicity of the trace is kept in $D$ dimensions
and 
\begin{equation}
\mathrm{Tr}(\gamma_5\overline{\gamma}^{\mu}\overline{\gamma}^{\nu}\overline{\gamma}^{\rho}\overline{\gamma}^{\sigma}) = -4i\varepsilon^{\mu\nu\rho\sigma}    
\end{equation}
is satisfied by the $4$-dimensional part of the $\gamma$-matrices.

In \texttt{FORM}, we implemented these algebraic relations in such a way 
that terms cancel as soon as possible in order to increase computational
performance at higher orders.
For instance, this is done by making use of the fact that
Lorentz index contractions between $4$-dimensional and evanescent 
quantities vanish, e.g.\
$\overline{\eta}_{\mu\nu}\mathrm{Tr}(\ldots\widehat{\gamma}^{\nu}\ldots)=0$.

Traces of $\gamma$-matrices require a focus on their own for our calculation.
Although \texttt{FORM} is equipped with highly efficient built-in routines 
to deal with $\gamma$-traces, problems arise
when dealing with $\gamma$-matrices in $D$ dimensions within the framework 
of the BMHV scheme.
On the one hand $D$-dimensional traces in \texttt{FORM} cannot deal with $\gamma_5$,
and on the other hand the BMHV-algebra presented above is not implemented in \texttt{FORM}.
We stress that traces of $4$-dimensional $\gamma$-traces in \texttt{FORM} are optimised using dedicated rules and relations, such as the so-called Chisholm identity, in order to keep the number of 
generated terms small, see Ref.\ \cite{FORM:Manual}.
These algorithms are highly efficient, well-established and have thoroughly been 
tested over the past years.
In order to make use of these optimisations as much as possible, we factorise
every $\gamma$-trace into a $D$-dimensional part that does not contain $\gamma_5$
and a $4$-dimensional part with possible occurrences of $\gamma_5$.

First, each $\gamma^{\mu}$-matrix is expressed via $\overline{\gamma}^{\mu}$
and $\widehat{\gamma}^{\mu}$. 
In the case of $D$-dimensional matrices, this is either done via
$\gamma^{\mu}=\overline{\gamma}^{\mu}+\widehat{\gamma}^{\mu}$ or,
slightly more efficient as it does not increase the number of terms, via
\begin{equation}
    \begin{aligned}
        \mathrm{Tr}(\ldots\mathbb{P}_{\alpha}\gamma^{\mu}\mathbb{P}_{\beta}\ldots)
        = 
        \begin{cases}
            \mathrm{Tr}(\ldots\overline{\gamma}^{\mu}\projR\ldots), &\alpha=\mathrm{L},\,\beta=\mathrm{R},\\
            \mathrm{Tr}(\ldots\overline{\gamma}^{\mu}\projL\ldots), &\alpha=\mathrm{R},\,\beta=\mathrm{L},\\
            \mathrm{Tr}(\ldots\widehat{\gamma}^{\mu}\projR\ldots), &\alpha=\mathrm{R},\,\beta=\mathrm{R},\\
            \mathrm{Tr}(\ldots\widehat{\gamma}^{\mu}\projL\ldots), &\alpha=\mathrm{L},\,\beta=\mathrm{L},\\
        \end{cases}
    \end{aligned}
\end{equation}
where $\mathbb{P}_\mathrm{L/R}=(\mathbb{1}\mp\gamma_5)/2$, 
depending on the considered situation.
Next, each $\gamma$-trace is brought into a ``normal form'',
where all evanescent matrices are on the left-most side inside the trace,
while all $4$-dimensional matrices are on the right-most side. 
This is achieved by applying the anticommutation relation 
$\overline{\gamma}^{\mu}\widehat{\gamma}^{\nu}=-\widehat{\gamma}^{\nu}\overline{\gamma}^{\mu}$,
see Eq.\ \eqref{Eq:BMHV-Algebra-1}.
More specifically, this normal form takes the form
\begin{equation}\label{Eq:TraceNormalForm}
    \begin{aligned}
        \mathrm{Tr}(\widehat{\gamma}^{\nu_1}\ldots\widehat{\gamma}^{\nu_m}\overline{\gamma}^{\mu_1}\ldots\overline{\gamma}^{\mu_n}\mathbb{\Lambda}),
    \end{aligned}
\end{equation}
with $\mathbb{\Lambda}\in\{\mathbb{1},\gamma_5,\projL,\projR\}$.
Clearly, such traces are non-vanishing only if $n$ and $m$  are even integers.
A trace of this form can then be factorised as
\begin{equation}\label{Eq:TraceFactorisationTheorem}
    \begin{aligned}
        \mathrm{Tr}(\widehat{\gamma}^{\nu_1}\ldots\widehat{\gamma}^{\nu_m}\overline{\gamma}^{\mu_1}\ldots\overline{\gamma}^{\mu_n}\mathbb{\Lambda})
        = \frac{1}{4} \mathrm{Tr}(\widehat{\gamma}^{\nu_1}\ldots\widehat{\gamma}^{\nu_m}) \mathrm{Tr}(\overline{\gamma}^{\mu_1}\ldots\overline{\gamma}^{\mu_n}\mathbb{\Lambda}).
    \end{aligned}
\end{equation}
A concise proof of this formula is provided in App.\ \ref{App:TraceFactorisation}.
In \texttt{FORM}, traces of the type 
$\mathrm{Tr}(\widehat{\gamma}^{\nu_1}\ldots\widehat{\gamma}^{\nu_m})$ 
can be evaluated with the built-in $D$-dimensional $\gamma$-trace algorithms, 
while traces of the type $\mathrm{Tr}(\overline{\gamma}^{\mu_1}\ldots\overline{\gamma}^{\mu_n}\mathbb{\Lambda})$
can be handled by the built-in $4$-dimensional $\gamma$-trace algorithms utilising
all algorithmic benefits.
For the latter, the occurrence of $\gamma_5$ does not result in any problems
as these traces are now handled in purely $4$ dimensions.
Clearly, the BMHV-algebra, as shown in Eqs.\ \eqref{Eq:BMHV-Metric-Contractions}
to \eqref{Eq:BMHV-Algebra-2}, must be implemented in \texttt{FORM} 
by hand and every step towards Eq.\ \eqref{Eq:TraceFactorisationTheorem} must respect it.

\subsection{Tadpole Decomposition}\label{Sec:TadpoleDecomposition}

As mentioned above, in order to extract UV divergences,
we employ a tadpole decomposition following the approach of Refs.\
\cite{Misiak:1994zw,Chetyrkin:1997fm}.
This method is based on recursively applying
the exact identity
\begin{equation} \label{Eq:StandardTadpoleExpansion}
    \begin{aligned}
        \frac{1}{(k+p)^2} = \frac{1}{k^2 - M^2} - \frac{p^2 + 2 \, k \cdot p + M^2}{k^2 - M^2} \, \frac{1}{(k+p)^2}
    \end{aligned}
\end{equation}
to the propagators,
where $k$ denotes a loop momentum or any linear combination thereof,
and $M$ is an auxiliary mass scale.
The expansion is performed to an order sufficient to capture all UV-divergent
contributions. 
%as determined by the degree of divergence.

While Eq.\ \eqref{Eq:StandardTadpoleExpansion} is exact, its direct use 
in the context of the tadpole decomposition proves to be inconvenient 
in automated multi-loop computations.
In particular, higher order applications face challenges related 
to a consistent momentum routing across genuine $L$-loop 
and associated $(<\!L)$-loop counterterm-inserted
Feynman diagrams, due to subdivergences.
To circumvent these difficulties, we employ
an improved tadpole expansion, as suggested in 
Refs.\ \cite{Misiak:1994zw,Chetyrkin:1997fm} and elaborated in
detail in Ref.\ \cite{Lang:2020nnl} (see also Ref.\ \cite{Stockinger:2023ndm}).
In this approach, the auxiliary mass parameter $M$ is introduced 
uniformly in all propagators, and a Taylor-expansion in external momenta 
(and internal/physical masses, if present) is performed.
The expansion is truncated at the order determined by the 
superficial degree of divergence of the considered Green function. 
For example, for a massless propagator $(k+p)^{-2}$, we obtain
\begin{equation}\label{Eq:TadpoleDecomposition}
    \begin{aligned}
        %\frac{1}{(k+p)^2} \longrightarrow \frac{1}{(k+p)^2 - M^2}
        \frac{1}{(k+p)^2 - M^2}
        = \, \frac{1}{k^2 - M^2} 
        &- \frac{2 \, k \cdot p}{(k^2 - M^2)^2} 
        - \frac{p^2}{(k^2 - M^2)^2} + \frac{4 (k \cdot p)^2}{(k^2 - M^2)^3}\\  
        &+ \frac{4 \, p^2 \, k \cdot p}{(k^2 - M^2)^3} 
        - \frac{8 (k \cdot p)^3}{(k^2 - M^2)^4}  
        + \mathcal{O}\big(p^4\big).
    \end{aligned}
\end{equation}
Specifically, for the Green function $cB^{\mu}$, 
which has degree of divergence $3$ 
(see Fig.\ \ref{Fig:cA-4-Loop-Diagrams} for representative diagrams),
the expansion must be carried out up to third order, 
as shown in Eq.\ \eqref{Eq:TadpoleDecomposition}.
In contrast, for the Green function $cB^{\mu}B^{\nu}B^{\rho}$, 
whose degree of divergence is $1$ 
(see Fig.\ \ref{Fig:cAAA-4-Loop-Diagrams} and \ref{Fig:cAAA-3LCT1-Diagrams} 
for representative diagrams),
the expansion can be truncated after the second term on the 
RHS of Eq.\ \eqref{Eq:TadpoleDecomposition}.
In the case of the Green function $B^{\mu}B^{\nu}B^{\rho}B^{\sigma}$, 
only the zeroth-order term --- independent of external momenta ---
must be retained, as its degree of divergence is $0$. 

Rearranging the terms on the RHS of Eq.\ \eqref{Eq:TadpoleDecomposition} yields
\begin{equation}\label{Eq:TadpoleDecomposition-Rearranged}
    \begin{aligned}
        %\frac{1}{(k+p)^2} \longrightarrow \frac{1}{(k+p)^2 - M^2} = 
        \frac{1}{k^2 - M^2} - \frac{p^2 + 2 \, k \cdot p}{(k^2 - M^2)^2}  + \frac{(p^2 + 2 \, k \cdot p)^2}{(k^2 - M^2)^3} + \ldots,
    \end{aligned}
\end{equation}
which resembles the recursive application of the exact tadpole decomposition 
in Eq.\ (\ref{Eq:StandardTadpoleExpansion}) 
up to 
terms $\propto M^2$ in the numerator.
In the Taylor-expansion approach, omitting such numerator terms $\propto M^2$
necessitates a systematic compensation at each order, particularly 
at the multi-loop level in the presence of subdivergences.
This is achieved by including all required auxiliary counterterms proportional to $M^2$
in order to subtract the effect of the missing numerator terms $\propto M^2$.
Both the auxiliary mass parameter $M^2$ and the auxiliary mass counterterms 
are used only for technical purposes in the evaluation of the Feynman integrals, and
are not intrinsic to the theory.
In particular, the auxiliary mass counterterm associated to the gauge boson 
does not break gauge invariance, as it does not enter the action.

To some extent, this method addresses both issues outlined 
in Sec.\ \ref{Sec:BMHVSpecificChallenges}
by decomposing all integrals into tadpoles without external momenta,
thereby reducing the computational complexity in certain aspects.
This allows the renormalisation of chiral gauge theories
in the BMHV scheme at higher loop orders, where Green functions with 
up to five external legs must be computed. 
Its main drawback, however, is the proliferation of terms 
that arises after decomposing the integrand according 
to Eq.\ \eqref{Eq:TadpoleDecomposition},
particularly for Green functions with a high degree of divergence, such as $cB^{\mu}$.

\subsection{Practical Implementation of Counterterms}

In practice, we implemented counterterm insertions into propagators in \texttt{FORM} 
by employing modified propagator expressions, which are applied recursively depending 
on the required insertion order.
For fermions, we have
\begin{equation}\label{Eq:RecursiveFermionPropagator}
    \begin{aligned}
        \mathcal{D}_{f,ij}(k) &= \frac{i \slashed{k}}{k^2} \bigg( \delta_{ij} 
        + i
        \!\!\!\!\!\!\sum_{\alpha,\beta\in\{\mathrm{L},\mathrm{R}\}}\!\!\!\!\!\!\delta Z^{f}_{\overline{\alpha}\beta,ik} \,
        \mathbb{P}_{\alpha} \slashed{k} \mathbb{P}_{\beta} \,
        \mathcal{D}_{f,kj}(k) \bigg),
    \end{aligned}
\end{equation}
where an index with a bar, e.g. $\overline{\alpha}$, denotes the opposite chirality
of the actual index, i.e. $\overline{\alpha}=\mathrm{R}$ for $\alpha=\mathrm{L}$.
For the gauge boson, the propagator takes the form
\begin{equation}\label{Eq:RecursiveGaugeBosonPropagator}
    \begin{aligned}
        \mathcal{D}_g^{\mu\nu}(k) = \frac{-i}{k^2} \Big(\eta^{\mu\rho} &- \big(1 - \xi\big) \frac{k^{\mu}k^{\rho}}{k^2}\Big) 
        \Big( \delta_{\rho}^{\phantom{\rho}\nu} - i \Big[ 
        - \delta\overline{Z}^{g}_{M} M^2 \overline{\eta}_{\rho\sigma} 
        - \delta\widehat{Z}^{g}_{M} M^2 \widehat{\eta}_{\rho\sigma}\\
        &+ \delta Z^g_D \big( \eta_{\rho\sigma}k^2 - k_{\rho}k_{\sigma} \big) 
        + \delta \overline{Z}^g_{4D} \big( \overline{\eta}_{\rho\sigma}\overline{k}^2 - \overline{k}_{\rho}\overline{k}_{\sigma} \big)\\
        &+ \delta \overline{Z}^g_{11} \, \overline{\eta}_{\rho\sigma}\overline{k}^2
        + \delta \widehat{Z}^g_{12} \, \overline{\eta}_{\rho\sigma}\widehat{k}^2
        + \delta \widehat{Z}^g_{21} \, \widehat{\eta}_{\rho\sigma}\overline{k}^2
        + \delta \widehat{Z}^g_{22} \, \widehat{\eta}_{\rho\sigma}\widehat{k}^2\\
        &- \delta \overline{Z}^g_{33} \, \overline{k}_{\rho}\overline{k}_{\sigma}
        - \delta \widehat{Z}^g_{34} \, \overline{k}_{\rho}\widehat{k}_{\sigma}
        - \delta \widehat{Z}^g_{43} \, \widehat{k}_{\rho}\overline{k}_{\sigma}
        - \delta \widehat{Z}^g_{44} \, \widehat{k}_{\rho}\widehat{k}_{\sigma}
        \Big] \mathcal{D}_g^{\sigma\nu}(k) \Big).
    \end{aligned}
\end{equation}
This approach is similar to that of Ref.\ \cite{Luthe:2017ttg},
but adapted to our calculation in the BMHV scheme.
In particular, the counterterm contributions in the propagators
include all possible Lorentz covariants.
For each $\delta Z$, a perturbative expansion is implied.\footnote{The $\delta Z$'s do not 
generally result from a multiplicative renormalisation transformation,
as they also account for symmetry-restoring counterterms.
}
In the case of the gauge boson propagator, the first and the last two lines of 
Eq.\ (\ref{Eq:RecursiveGaugeBosonPropagator}) already form a complete basis.
For convenience, we also include the second line, which introduces both 
$D$- and $4$-dimensional transversal combinations
that frequently appear in practical applications, albeit with some redundancy.
In non-Abelian gauge theories with interacting ghosts or
theories with scalar bosons, analogous propagators have to
be implemented for the ghosts and the scalars, respectively.

Counterterm insertions for interaction vertices are implemented straightforwardly.
For instance, the gauge interaction is given by
\begin{equation}\label{Eq:GaugeInteraction-Implementation}
    \begin{aligned}
        - i g \!\!\!\!\!\!\sum_{\alpha,\beta\in\{\mathrm{L},\mathrm{R}\}}\!\!\!\!\! \big( \mathcal{Y}_{\overline{\alpha}\beta,ij} 
        + \delta \mathcal{Y}_{\overline{\alpha}\beta,ij} \big)
        \mathbb{P}_{\alpha} \gamma^{\mu} \mathbb{P}_{\beta}.
    \end{aligned}
\end{equation}
This expression represents a general implementation that includes both left- 
and right-handed fermions, as well as evanescent gauge interactions.

In contrast to the theory discussed in Ref.\ \cite{Ebert:2024xpy}, 
the right-handed model considered in this work, 
see Sec.\ \ref{Sec:TheoryDefinition},
has only nonzero $\mathcal{Y}_{RR}\equiv\hypR$, 
while all other hypercharge matrices vanish. 
Consequently, left-handed and evanescent gauge interactions are absent,
which in turn reduces the number of required counterterms.
Accordingly, from the subset of counterterms listed above, we specifically
need $\delta Z^{f}_{RR}$, $\delta\overline{Z}^{g}_{M}$, $\delta\overline{Z}^{g}_{4D}$,
$\delta\overline{Z}^{g}_{11}$, $\delta\widehat{Z}^{g}_{12}$ and $\delta \mathcal{Y}_{RR}$.
In Sec.~\ref{Sec:4-Loop-Renormalisation} we will decompose these renormalisation constants further into symmetric contributions $\delta Z$, non-symmetric contributions with $1/\epsilon$-poles $\delta X$, and finite symmetry-restoring counterterms $\mathcal{F}$.
With this convention, the corresponding 4-loop 
renormalisation constants in the model of Sec.\ \ref{Sec:TheoryDefinition} take the form
\begin{equation}
    \begin{aligned}
        \delta Z^{f,(4)}_{RR,ij} &= \frac{g^{8}}{(16 \pi^2)^4} \Big( \delta Z^{(4)}_{\psi,ij} + \delta \overline{X}^{(4)}_{\overline{\psi}\psi,ij} + \mathcal{F}_{\psi\overline{\psi}, ij}^{4,\text{break}} \Big),\\
        \delta \overline{Z}^{g,(4)}_{4D} &= \frac{g^{8}}{(16 \pi^2)^4} \, \delta Z^{(4)}_{B},\\ 
        \delta \overline{Z}^{g,(4)}_{11} &= \frac{g^{8}}{(16 \pi^2)^4} \Big( \delta \overline{X}^{(4)}_{BB} + \mathcal{F}_{BB}^{4,\text{break}} \Big),\\ 
        \delta \widehat{Z}^{g,(4)}_{12} &= \frac{g^{8}}{(16 \pi^2)^4} \, \delta \widehat{X}^{(4)}_{BB},\\ 
        \delta \mathcal{Y}^{(4)}_{RR,ij} &= \frac{g^{8}}{(16 \pi^2)^4} \, \big(\mathcal{Y}_R\big)_{ik} \delta Z^{(4)}_{\psi,kj},
    \end{aligned}
\end{equation}
with explicit expressions provided in Sec.\ \ref{Sec:4-Loop-Renormalisation}.
In addition, although not displayed explicitly here, symmetry-restoring quartic gauge 
boson counterterms $\delta \overline{Z}_{4g}$
and quantum corrections to the $\Delta$-operator are also required for the full renormalisation procedure,
as discussed in Sec.\ \ref{Sec:4-Loop-Renormalisation}.

\subsection{Tensor Reduction}\label{Sec:TensorReduction}

Calculations in gauge theories lead to tensor integrals $I^{\mu_1\cdots\mu_r}$ of rank $r$,
which we map to scalar integrals using tensor reduction.
One starts by writing the tensor integral
as a linear combination of all possible Lorentz tensors 
$T_a^{\mu_1\cdots\mu_r}$ allowed by symmetries:
\begin{equation}\label{Eq:GeneralTensorReduction}
    \begin{aligned}
        I^{\mu_1\cdots\mu_r}(p_1,\ldots,p_E) = 
        \sum_{a=1}^{n} C_a(p_1,\ldots,p_E) \, T_{a}^{\mu_1\cdots\mu_r}(p_1,\ldots,p_E).
    \end{aligned}
\end{equation}
The tensors $T_a^{\mu_1\cdots\mu_r}$ are independent of 
the perturbative order and the coefficients $C_a$ consist of scalar loop integrals.

After applying the tadpole decomposition, as discussed in Sec.\ \ref{Sec:TadpoleDecomposition}, the resulting
vacuum bubble tensor integrals do not depend on external momenta $p_i$.
Thus, both the tensors $T_a^{\mu_1\cdots\mu_r}$ and the
coefficients $C_a$ are independent of external momenta.
For this reason, the tensors $T_a^{\mu_1\cdots\mu_r}$ are solely composed
of products of metric tensors $\eta^{\mu\nu}$.
These metric tensors are $D$-dimensional, resulting in 
purely $D$-dimensional projectors, 
in line with the discussion in Sec.\ \ref{Sec:BMHVSpecificChallenges}.

In this work, we employ the method proposed 
and applied in Refs.\ \cite{Herzog:2017ohr,Ruijl:2018poj} 
for an efficient tensor reduction
of such vacuum bubble tensor integrals.
During the course of this project, two other
groups published work on tensor reduction, including
external momenta and providing further mathematical insights, 
see Refs.\ \cite{Anastasiou:2023koq} and \cite{Goode:2024mci}.
Notably, the group of Ref.\ \cite{Goode:2024mci} provided 
a detailed discussion of the method employed here
--- naming it \textit{orbit partition approach} ---
and published the tensor reduction software tool \texttt{OPITeR} \cite{Goode:2024cfy}.
Nonetheless, we find it valuable to document our implementation
of this method for the reduction of vacuum bubble tensor integrals,
specifically because all computations in this project have been performed
using our own \texttt{FORM} programs.

In a nutshell, the tensor reduction consists of two main tasks.
The first task is the construction of a basis $\mathcal{B}$ of all tensor structures
$T_a^{\mu_1\cdots\mu_r}$ contributing to a general tensor, see Eq.\ \eqref{Eq:GeneralTensorReduction},
and the determination of projectors $P_a^{\mu_1\cdots\mu_r}$ to calculate the coefficients $C_a$.
The projectors (dual tensors) are constructed from the basic tensors $T_a^{\mu_1\cdots\mu_r}$ as well and form the dual basis $\mathcal{B}^{*}$.
They are required to satisfy the orthonormality relation
\begin{equation}\label{Eq:OrthogonalityRelation}
    \begin{aligned}
        P_a^{\mu_1\cdots\mu_r} \, T_{b,\mu_1\cdots\mu_r} = \delta_{ab}\,.
    \end{aligned}
\end{equation}
This first task can be done once and in advance of the calculation of Green functions.
We collect the resulting projectors in so-called tensor reduction tables.
The second task is the application of the projectors to the specific tensor integrals of interest to obtain explicit expressions for the coefficients $C_a$.
Condition \eqref{Eq:OrthogonalityRelation} implies that the coefficients $C_a$ are obtained upon projection via
\begin{equation}\label{Eq:ProjectionOfCoefficients}
    \begin{aligned}
        C_a = P_a^{\mu_1\cdots\mu_r} \, I_{\mu_1\cdots\mu_r}
    \end{aligned}
\end{equation}
for a given loop integral $I_{\mu_1\cdots\mu_r}$.

\paragraph{Constructing the tensor reduction tables:}

In the present case of vacuum bubbles, only tensor integrals of even rank
$r$ give rise to non-vanishing contributions and, as mentioned above,
the only possible Lorentz covariants are metric tensors $\eta^{\mu\nu}$.
Hence, following Eq.\ 
(\ref{Eq:GeneralTensorReduction}), these vacuum bubble tensor integrals 
can be decomposed into a linear combination of $n=|S_{2}^{r}|=(r-1)!!=r!/(2^{r/2}(r/2)!)$ 
independent tensors
\begin{equation}\label{Eq:GeneralTensorStructureForVacuumBubbles}
    \begin{aligned}
        T_{\sigma}^{\mu_1\cdots\mu_r} = \eta^{\mu_{\sigma(1)}\mu_{\sigma(2)}} \cdots \eta^{\mu_{\sigma(r-1)}\mu_{\sigma(r)}},
    \end{aligned}
\end{equation}
composed of $r/2$ metric tensors each.
$S_2^r$ is the set of permutations that generate all distinct products of
metric tensors of rank $r$, i.e.\ all $T_a^{\mu_1\cdots\mu_r}$.

Clearly, the projectors can only consist of such products of metric tensors as well.
Therefore, with all possible tensor structures $T_a^{\mu_1\cdots\mu_r}$ at hand,
%see Eq.\ \eqref{Eq:GeneralTensorStructureForVacuumBubbles}, 
the projectors 
are generally given as a linear combination of these tensor structures as 
\begin{equation}\label{Eq:GeneralProjectors}
    \begin{aligned}
        P_a^{\mu_1\cdots\mu_r} = \sum_{b=1}^{n} A^{(a)}_{b} \, T_b^{\mu_1\cdots\mu_r},
    \end{aligned}
\end{equation}
with yet unknown coefficients $A^{(a)}_b$.

Using the orthonormality relation (\ref{Eq:OrthogonalityRelation})
for a fixed $a\in\{1,\ldots,n\}$,
we obtain $n$ equations for $n$ coefficients $A^{(a)}_b$ upon 
contraction with the $n$ tensors $\{T_{b}^{\mu_1\cdots\mu_r}\}_{b=1}^{n}=\mathcal{B}$.
Solving this system of equations determines the projector $P_a^{\mu_1\cdots\mu_r}$.
Indeed, in this way all projectors $P_a^{\mu_1\cdots\mu_r}$, $\forall\,a\in\{1,\ldots,n\}$,
are determined, since all other projectors (with $b\neq a$) can be obtained by permutations of Lorentz indices.\footnote{
Given $\sigma \in S_r$, such that $\sigma \circ T_a^{\mu_1\cdots\mu_r} = T_b^{\mu_1\cdots\mu_r}$, then $P_b^{\mu_1\cdots\mu_r} = \sigma \circ P_a^{\mu_1\cdots\mu_r}$.
}
However, solving this system of equations means inverting an $n \times n$ matrix.
This becomes a major challenge as $n$ is growing very 
fast with the tensor rank $r$, see Tab.\ \ref{Tab:TensorSize}.
\begin{table}[t]
    \centering
    \begin{tabular}{|c||c|c|c|c|c|c|c|c|} \hline 
        $r$ & 2 & 4 & 6 & 8 & 10 & 12 & 14 & 16\\ \hline 
        $n=|S_{2}^{r}|$ & 1 & 3 & 15 & 105 & 945 & 10395 & 135135 & 2027025\\ \hline 
        $m=p(r/2)$ & 1 & 2 & 3 & 5 & 7 & 11 & 15 & 22\\ \hline
    \end{tabular}
    \caption{Numbers $n$ and $m$ of independent tensor structures of rank $r$ vacuum tensor integrals and integer partitions of $r/2$, respectively, for $r\leq16$.}
    \label{Tab:TensorSize}
\end{table}

This problem can be approached using symmetry arguments, 
as explained in Refs.\ \cite{Herzog:2017ohr,Ruijl:2018poj,Anastasiou:2023koq,Goode:2024mci},
which effectively reduce the size of the system of equations significantly.
In particular, for a rank $r$ tensor integral, 
the projector $P_a^{\mu_1\cdots\mu_r}$
is invariant under the same (index permutation) symmetries as the tensor 
$T_{a}^{\mu_1\cdots\mu_r}$, given by the corresponding 
\emph{stabiliser subgroup} $H(T_a)$.
For fixed $a\in\{1,\ldots,n\}$, all tensors $\{T_{b}^{\mu_1\cdots\mu_r}\}_{b=1}^{n}$ can be grouped into
$m$ disjoint \emph{orbits} $\{\tensor*[_r]{\Theta}{_c^{(a)}}\}_{c=1}^{m}$, 
each consisting of tensors that transform into one another under the action of $H(T_a)$, 
forming a partition of $\mathcal{B}$.
For each $c\in\{1,\ldots,m\}$, the sum of all elements of 
$\tensor*[_r]{\Theta}{_c^{(a)}}$
resembles a tensor that has the same (permutation) symmetries as
$T_{a}^{\mu_1\cdots\mu_r}$, i.e.\ is invariant under
$H(T_a)$.
Due to these symmetry properties, all tensors of a given orbit
come with the same coefficient $A_c^{(a)}$.
Thus, the RHS of Eq.\ \eqref{Eq:GeneralProjectors} can be 
rearranged to
\begin{equation}\label{Eq:MinimalProjectors}
    \begin{aligned}
        P_a^{\mu_1\cdots\mu_r} = \sum_{c=1}^{m} A^{(a)}_{c} \Bigg( \sum_{ T \in \tensor*[_r]{\Theta}{_c^{(a)}} } T^{\mu_1\cdots\mu_r} \Bigg),
    \end{aligned}
\end{equation}
with $m$ coefficients $A^{(a)}_{c}$.
Since each bracketed sum of tensors on the RHS of Eq.\
\eqref{Eq:MinimalProjectors} is stabilised under $H(T_a)$ by construction,
this Ansatz automatically ensures the correct symmetry properties.

The number of orbits $m$ depends on the rank $r$ of the problem and
is determined by the number of the possible index-contraction \emph{cycles}.
These \emph{cycles} are closed contraction loops formed between the 
Lorentz indices of the metric tensors in $T_a^{\mu_1\cdots\mu_r}$
and those in a considered tensor $T_{b,\mu_1\cdots\mu_r}$.
For fixed $a$, tensors $T_{b,\mu_1\cdots\mu_r}$ that belong to the same 
orbit $\tensor*[_r]{\Theta}{_c^{(a)}}$
can thus be identified by the respective number of cycles.
The number of maximally possible cycles is equal to the number of integer partitions of $r/2$,
and hence $m$ grows significantly slower than $n$,
as shown in Tab.\ \ref{Tab:TensorSize}.
Consequently, for fixed $a\in\{1,\ldots,n\}$, contracting the projector 
$P_a^{\mu_1\cdots\mu_r}$ with a single representative tensor from each 
of the $m$ orbits $\tensor*[_r]{\Theta}{_c^{(a)}}$ and
using the orthonormality relation \eqref{Eq:OrthogonalityRelation}
results in only $m$ equations for $m$ coefficients $A^{(a)}_{c}$.
Solving this system of equations constitutes a significant simplification
compared to the original one of $n$ equations, cf.\ Tab.\ \ref{Tab:TensorSize}.
We label the orbits $\{\tensor*[_r]{\Theta}{_c^{(a)}}\}_{c=1}^{m}$ such that the orbit with $c=1$ contains the tensors exhibiting the maximal number of index-contraction cycles with $T_{a}^{\mu_1\cdots\mu_r}$, and the $c=m$ orbit those with the minimal number, i.e.\ a single cycle.
Notably, the orbit with $c=1$ consists of only one tensor, the
tensor $T_{a}^{\mu_1\cdots\mu_r}$ itself.
The construction of these orbits via contraction cycles is best illustrated with an explicit example.

In the following, let us consider a tensor integral problem of rank $r=6$:
After identifying all $n=15$ independent tensors, 
$\mathcal{B}=\{\sigma \circ T^{\mu_1\cdots\mu_6} \, | \, \sigma \in S_{2}^{6}\}$,
cf.\ Eq.\ \eqref{Eq:GeneralTensorStructureForVacuumBubbles}, we can, w.l.o.g., 
focus on fixed $a=1$ and construct $P_1^{\mu_1\cdots\mu_6}$ according to
Eq.\ \eqref{Eq:MinimalProjectors}.
To achieve this, we analyse the Lorentz index
contractions of
\begin{equation}
T_1^{\mu_1\cdots\mu_6}=\eta^{\mu_{1}\mu_{2}}\eta^{\mu_{3}\mu_{4}}\eta^{\mu_{5}\mu_{6}}
\end{equation}
with $\{T_{b,\mu_1\cdots\mu_6}\}_{b=1}^{15}$
in order to identify all orbits 
$\{\tensor*[_6]{\Theta}{_c^{(1)}}\}_{c=1}^{m}$ (with yet unknown $m$).
Indeed, in the present case of $r=6$, 
we can identify a maximum of $m=3$ disjoint cycles of index contractions, 
which is exactly the number of integer partitions of $r/2$.
As mentioned above, the first orbit $\tensor*[_6]{\Theta}{_1^{(1)}}$ 
consists solely of $T_{1,\mu_1\cdots\mu_6}$,
since it is the only tensor that can be reached by acting with $h \in H(T_1)$ on
itself, and thus inherently possesses already on its own the same permutation
symmetry as $T_1^{\mu_1\cdots\mu_6}$.
Moreover, because of this, it can be identified as the only tensor which admits 
the most possible contraction cycles with $T_1^{\mu_1\cdots\mu_6}$, provided by
\begin{equation}\label{Eq:ContractionCycles-3}
    \begin{aligned}
        \contraction[1.25ex]{ \!\! }{ \eta^{\mu_{1}\mu_{2}} }{ \eta^{\mu_{3}\mu_{4}} \eta^{\mu_{5}\mu_{6}} \, }{ \eta_{\mu_{1}\mu_{2}} }
        \bcontraction[1.25ex]{ \,\,\, }{ \eta^{\mu_{1}\mu_{2}} }{ \eta^{\mu_{3}\mu_{4}} \eta^{\mu_{5}\mu_{6}} \, }{ \eta_{\mu_{1}\mu_{2}} }
        \contraction[2ex]{ \eta^{\mu_{1}\mu_{2}} \!\! }{ \eta^{\mu_{3}\mu_{4}} }{ \eta^{\mu_{5}\mu_{6}} \, \eta_{\mu_{1}\mu_{2}} }{ \eta_{\mu_{3}\mu_{4}} }
        \bcontraction[2ex]{ \eta^{\mu_{1}\mu_{2}} \,\,\, }{ \eta^{\mu_{3}\mu_{4}} }{ \eta^{\mu_{5}\mu_{6}} \, \eta_{\mu_{1}\mu_{2}} }{ \eta_{\mu_{3}\mu_{4}} }
        \contraction[2.75ex]{ \eta^{\mu_{1}\mu_{2}} \eta^{\mu_{3}\mu_{4}} \!\! }{ \eta^{\mu_{5}\mu_{6}} }{ \, \eta_{\mu_{1}\mu_{2}} \eta_{\mu_{3}\mu_{4}} }{ \eta_{\mu_{5}\mu_{6}} }
        \bcontraction[2.75ex]{ \eta^{\mu_{1}\mu_{2}} \eta^{\mu_{3}\mu_{4}} \,\,\, }{ \eta^{\mu_{5}\mu_{6}} }{ \, \eta_{\mu_{1}\mu_{2}} \eta_{\mu_{3}\mu_{4}} }{ \eta_{\mu_{5}\mu_{6}} }
        \eta^{\mu_{1}\mu_{2}} \eta^{\mu_{3}\mu_{4}} \eta^{\mu_{5}\mu_{6}} \, \eta_{\mu_{1}\mu_{2}} \eta_{\mu_{3}\mu_{4}} \eta_{\mu_{5}\mu_{6}}.
    \end{aligned}
\end{equation}
In particular, these cycles are given by 
$1\rightarrow2\rightarrow1$, $3\rightarrow4\rightarrow3$ and $5\rightarrow6\rightarrow5$,
fixing $m=3$ in the case of $r=6$.
Hence, for the first orbit, we find 
\begin{equation}\label{Eq:Rank6-SymmetrySet-1}
    \begin{aligned}
        \tensor*[_6]{\Theta}{_1^{(1)}} \equiv \{T_{1,\mu_1\cdots\mu_6} = \eta_{\mu_{1}\mu_{2}} \eta_{\mu_{3}\mu_{4}} \eta_{\mu_{5}\mu_{6}}\}.
    \end{aligned}
\end{equation}
Next, the second orbit, $\tensor*[_6]{\Theta}{_2^{(1)}}$,
is filled with all tensors
that admit two contraction cycles with $T_1^{\mu_1\cdots\mu_6}$.
For instance, this is the case for the tensor 
$T_{4,\mu_1\cdots\mu_6}=\eta_{\mu_{1}\mu_{3}}\eta_{\mu_{2}\mu_{4}}\eta_{\mu_{5}\mu_{6}}$, 
where we find
\begin{equation}\label{Eq:ContractionCycles-2}
    \begin{aligned}
        \contraction[1.25ex]{ \!\! }{ \eta^{\mu_{1}\mu_{2}} }{ \eta^{\mu_{3}\mu_{4}} \eta^{\mu_{5}\mu_{6}} \, }{ \eta_{\mu_{1}\mu_{3}} }
        \bcontraction[2ex]{ \,\,\, }{ \eta^{\mu_{1}\mu_{2}} }{ \eta^{\mu_{3}\mu_{4}} \eta^{\mu_{5}\mu_{6}} \, \eta_{\mu_{1}\mu_{3}} }{ \!\!\!\!\!\!\!\!\!\! \eta_{\mu_{2}\mu_{4}} }
        \bcontraction[1.25ex]{ \eta^{\mu_{1}\mu_{2}} \!\! }{ \eta^{\mu_{3}\mu_{4}} }{ \eta^{\mu_{5}\mu_{6}} \, }{ \quad\,\,\, \eta_{\mu_{1}\mu_{3}} }
        \contraction[2ex]{ \eta^{\mu_{1}\mu_{2}} \,\,\, }{ \eta^{\mu_{3}\mu_{4}} }{ \eta^{\mu_{5}\mu_{6}} \, \eta_{\mu_{1}\mu_{3}} }{ \eta_{\mu_{2}\mu_{4}} }
        \contraction[2.75ex]{ \eta^{\mu_{1}\mu_{2}} \eta^{\mu_{3}\mu_{4}} \!\! }{ \eta^{\mu_{5}\mu_{6}} }{ \, \eta_{\mu_{1}\mu_{2}} \eta_{\mu_{3}\mu_{4}} }{ \eta_{\mu_{5}\mu_{6}} }
        \bcontraction[2.75ex]{ \eta^{\mu_{1}\mu_{2}} \eta^{\mu_{3}\mu_{4}} \,\,\, }{ \eta^{\mu_{5}\mu_{6}} }{ \, \eta_{\mu_{1}\mu_{2}} \eta_{\mu_{3}\mu_{4}} }{ \eta_{\mu_{5}\mu_{6}} }
        \eta^{\mu_{1}\mu_{2}} \eta^{\mu_{3}\mu_{4}} \eta^{\mu_{5}\mu_{6}} \, \eta_{\mu_{1}\mu_{3}} \eta_{\mu_{2}\mu_{4}} \eta_{\mu_{5}\mu_{6}},
    \end{aligned}
\end{equation}
i.e.\ the two cycles $1\rightarrow3\rightarrow4\rightarrow2\rightarrow1$ and $5\rightarrow6\rightarrow5$, and  analogously for all other 
$T_{b,\mu_1\cdots\mu_6}\in\tensor*[_6]{\Theta}{_2^{(1)}}$.
Thus, we have
\begin{equation}\label{Eq:Rank6-SymmetrySet-2}
    \begin{aligned}
        \tensor*[_6]{\Theta}{_2^{(1)}}\equiv
        \{
        &\eta_{\mu_{1}\mu_{2}} \eta_{\mu_{3}\mu_{5}} \eta_{\mu_{4}\mu_{6}},
        \eta_{\mu_{1}\mu_{2}} \eta_{\mu_{3}\mu_{6}} \eta_{\mu_{4}\mu_{5}},
        \eta_{\mu_{1}\mu_{3}} \eta_{\mu_{2}\mu_{4}} \eta_{\mu_{5}\mu_{6}}\\
        &\eta_{\mu_{1}\mu_{4}} \eta_{\mu_{2}\mu_{3}} \eta_{\mu_{5}\mu_{6}},
        \eta_{\mu_{1}\mu_{5}} \eta_{\mu_{2}\mu_{6}} \eta_{\mu_{3}\mu_{4}},
        \eta_{\mu_{1}\mu_{6}} \eta_{\mu_{2}\mu_{5}} \eta_{\mu_{3}\mu_{4}}
        \}.
    \end{aligned}
\end{equation}
Finally, the last orbit $\tensor*[_6]{\Theta}{_3^{(1)}}$ is given by all
tensors that have only one index-contraction cycle with 
$T_1^{\mu_1\cdots\mu_6}$. 
An example for this set is provided by 
$T_{10,\mu_1\cdots\mu_6}=\eta_{\mu_{1}\mu_{5}}\eta_{\mu_{2}\mu_{3}}\eta_{\mu_{4}\mu_{6}}$.
In this case, we find
\begin{equation}\label{Eq:ContractionCycles-1}
    \begin{aligned}
        \contraction[1.25ex]{ \!\! }{ \eta^{\mu_{1}\mu_{2}} }{ \eta^{\mu_{3}\mu_{4}} \eta^{\mu_{5}\mu_{6}} \, }{ \eta_{\mu_{1}\mu_{5}} }
        \bcontraction[2.75ex]{ \,\,\, }{ \eta^{\mu_{1}\mu_{2}} }{ \eta^{\mu_{3}\mu_{4}} \eta^{\mu_{5}\mu_{6}} \, \eta_{\mu_{1}\mu_{5}} }{ \!\!\!\!\!\!\!\!\!\! \eta_{\mu_{2}\mu_{3}} }
        \contraction[2.75ex]{ \eta^{\mu_{1}\mu_{2}} \!\! }{ \eta^{\mu_{3}\mu_{4}} }{ \eta^{\mu_{5}\mu_{6}} \, \eta_{\mu_{1}\mu_{5}} }{ \quad\,\,\, \eta_{\mu_{2}\mu_{3}} }
        \bcontraction[2ex]{ \eta^{\mu_{1}\mu_{2}} \,\,\, }{ \eta^{\mu_{3}\mu_{4}} }{ \eta^{\mu_{5}\mu_{6}} \, \eta_{\mu_{1}\mu_{5}} \eta_{\mu_{2}\mu_{3}} }{ \!\!\!\!\!\!\!\!\!\! \eta_{\mu_{4}\mu_{6}} }
        \bcontraction[1.25ex]{ \eta^{\mu_{1}\mu_{2}} \eta^{\mu_{3}\mu_{4}} \!\! }{ \eta^{\mu_{5}\mu_{6}} }{ \, }{ \quad\,\,\, \eta_{\mu_{1}\mu_{5}} }
        \contraction[2ex]{ \eta^{\mu_{1}\mu_{2}} \eta^{\mu_{3}\mu_{4}} \,\,\, }{ \eta^{\mu_{5}\mu_{6}} }{ \, \eta_{\mu_{1}\mu_{5}} \eta_{\mu_{2}\mu_{3}} }{ \eta_{\mu_{5}\mu_{6}} }
        \eta^{\mu_{1}\mu_{2}} \eta^{\mu_{3}\mu_{4}} \eta^{\mu_{5}\mu_{6}} \, \eta_{\mu_{1}\mu_{5}} \eta_{\mu_{2}\mu_{3}} \eta_{\mu_{4}\mu_{6}},
    \end{aligned}
\end{equation}
indeed, with only one cycle $1\rightarrow5\rightarrow6\rightarrow4\rightarrow3\rightarrow2\rightarrow1$.
Hence, the last orbit is found to be
\begin{equation}\label{Eq:Rank6-SymmetrySet-3}
    \begin{aligned}
        \tensor*[_6]{\Theta}{_3^{(1)}}\equiv
        \{
        &\eta_{\mu_{1}\mu_{3}} \eta_{\mu_{2}\mu_{5}} \eta_{\mu_{4}\mu_{6}},
        \eta_{\mu_{1}\mu_{3}} \eta_{\mu_{2}\mu_{6}} \eta_{\mu_{4}\mu_{5}},
        \eta_{\mu_{1}\mu_{4}} \eta_{\mu_{2}\mu_{5}} \eta_{\mu_{3}\mu_{6}},
        \eta_{\mu_{1}\mu_{4}} \eta_{\mu_{2}\mu_{6}} \eta_{\mu_{3}\mu_{5}},\\
        &\eta_{\mu_{1}\mu_{5}} \eta_{\mu_{2}\mu_{3}} \eta_{\mu_{4}\mu_{6}},
        \eta_{\mu_{1}\mu_{5}} \eta_{\mu_{2}\mu_{4}} \eta_{\mu_{3}\mu_{6}},
        \eta_{\mu_{1}\mu_{6}} \eta_{\mu_{2}\mu_{3}} \eta_{\mu_{4}\mu_{5}},
        \eta_{\mu_{1}\mu_{6}} \eta_{\mu_{2}\mu_{4}} \eta_{\mu_{3}\mu_{5}}
        \}.
    \end{aligned}
\end{equation}
Therefore, the projector $P_1^{\mu_1\cdots\mu_6}$ is given according to Eq.\
\eqref{Eq:MinimalProjectors} with $m=3$ different coefficients $\{A^{(1)}_c\}_{c=1}^3$ 
and tensors provided in the orbits \eqref{Eq:Rank6-SymmetrySet-1}, 
\eqref{Eq:Rank6-SymmetrySet-2} and \eqref{Eq:Rank6-SymmetrySet-3}.
Contracting the projector $P_1^{\mu_1\cdots\mu_6}$ with a representative tensor
of each orbit in Eqs.\ \eqref{Eq:Rank6-SymmetrySet-1}, 
\eqref{Eq:Rank6-SymmetrySet-2} and \eqref{Eq:Rank6-SymmetrySet-3},\footnote{
Here, we used the same tensors as in the demonstrations in Eqs.\
\eqref{Eq:ContractionCycles-3}, \eqref{Eq:ContractionCycles-2} and 
\eqref{Eq:ContractionCycles-1}, i.e.\
$T_1^{\mu_1\cdots\mu_6}$, $T_4^{\mu_1\cdots\mu_6}$ 
and $T_{10}^{\mu_1\cdots\mu_6}$, respectively.
}
and using the orthonormality relation \eqref{Eq:OrthogonalityRelation}, 
gives rise to the $3$ equations
\begin{equation}
    \begin{aligned}
        1 &= P_1^{\mu_1\cdots\mu_6} T_{1,\mu_1\cdots\mu_6}
           = D^3 A_1^{(1)} + 6 D^2 A_2^{(1)} + 8 D A_3^{(1)},\\
        0 &= P_1^{\mu_1\cdots\mu_6} T_{4,\mu_1\cdots\mu_6} 
           = D^2 A_1^{(1)} + D \big( D^2 + D + 4 \big) A_2^{(1)} + 4D \big( D + 1 \big) A_3^{(1)},\\
        0 &= P_1^{\mu_1\cdots\mu_6} T_{10,\mu_1\cdots\mu_6} 
           = D A_1^{(1)} + 3 D \big( D + 1 \big) A_2^{(1)} + D \big( D^2 + 3 D + 4 \big) A_3^{(1)}.
    \end{aligned}
\end{equation}
These equations lead to the three coefficients
\begin{equation}\label{Eq:ProjectorCoeffs-Rank6}
    \begin{aligned}
        A^{(1)}_1 &= \frac{D^2+3D-2}{(D^2+2D-8)(D^2+D-2)D},\\
        A^{(1)}_2 &= - \frac{1}{(D^2+2D-8)(D-1)D},\\
        A^{(1)}_3 &= \frac{2}{(D^2+2D-8)(D^2+D-2)D},
    \end{aligned}
\end{equation}
which determine the projector $P_1^{\mu_1\cdots\mu_6}$.
All other projectors $P_{b\neq1}^{\mu_1\cdots\mu_6}$ are then obtained 
by permutation of the Lorentz indices.
In \texttt{FORM}, this is implemented via so-called wildcards.

Following this approach for all other tensor ranks, we have precomputed
all coefficients and projectors up to tensor rank $r\leq16$.
These are written into tables and called during the execution of the
tensor reduction routine in the \texttt{FORM} program.

\paragraph{Applying the tensor reduction to the tensor integrals:}

In contrast to the tensor reduction tables, which can be precomputed once,
this task needs to be done for each tensor integral during the 
actual computation. 
It mainly amounts to determining the coefficients $C_a$
associated to the considered tensor integral, see Eq.\ \eqref{Eq:ProjectionOfCoefficients}.
Symmetry arguments can be used to optimise the performance
of this task as well.
In particular, the symmetry of the respective integrand of the 
tensor integral is used, which needs to remain unchanged upon tensor reduction.

In alignment with the example discussed above, we consider a
rank $r=6$ tensor integral of the form
\begin{equation}\label{Eq:Rank6-TensorIntegral}
    \begin{aligned}
        I^{\mu_1\cdots\mu_6} 
        &=
        \big(\mu^{4-D}\big)^L 
        \int \Bigg( \prod_{i=1}^{L} \frac{d^Dk_i}{(2\pi)^D} \Bigg) 
        \frac{k_1^{\mu_1}k_1^{\mu_2}k_2^{\mu_3}k_2^{\mu_4}k_2^{\mu_5}k_2^{\mu_6}}{\mathcal{D}_1^{\nu_1} \cdots \mathcal{D}_N^{\nu_N}}
    \end{aligned}
\end{equation}
to exemplarily illustrate the procedure.
Clearly, the integrand of Eq.\ \eqref{Eq:Rank6-TensorIntegral}
admits an index permutation symmetry of the
two Lorentz indices $\{\mu_1,\mu_2\}$ and the four
Lorentz indices $\{\mu_3,\mu_4,\mu_5,\mu_6\}$ among each other,
respectively.
Due to this symmetry, the tensor integral decomposition, see Eq.\ 
\eqref{Eq:GeneralTensorReduction}, reduces to
\begin{equation}\label{Eq:TensorReduced-Rank6-TensorIntegral}
    \begin{aligned}
        I^{\mu_1\cdots\mu_6} &= C_1 \, \big( T_1 + T_2 + T_3 \big)^{\mu_1\cdots\mu_6}
        + C_4 \, \big( T_4 + T_5 + T_6 + \ldots + T_{15} \big)^{\mu_1\cdots\mu_6},
    \end{aligned}
\end{equation}
with only two (instead of $n=15$) distinct coefficients,
such that each of the two summands is invariant under the same 
(index permutation) symmetry as the integrand in Eq.\ \eqref{Eq:Rank6-TensorIntegral}.
The two coefficients can be obtained upon projection, see Eq.\ \eqref{Eq:ProjectionOfCoefficients},
as\footnote{Using the results for $A^{(1)}_c$ in Eq.\ \eqref{Eq:ProjectorCoeffs-Rank6}.}
\begin{equation}
    \begin{aligned}
        C_1 &= P_1^{\mu_1\cdots\mu_6} \, I_{\mu_1\cdots\mu_6}
        =
        \int \Bigg( \prod_{i=1}^{L} \frac{d^Dk_i}{(2\pi)^D} \Bigg) 
        \frac{\big(\mu^{4-D}\big)^L}{\mathcal{D}_1^{\nu_1} \cdots \mathcal{D}_N^{\nu_N}} \frac{(D+3)k_1^2k_2^4-4(k_1\cdot k_2)^2k_2^2}{(D-1)D(D+2)(D+4)},\\
        C_4 &= P_4^{\mu_1\cdots\mu_6} \, I_{\mu_1\cdots\mu_6}
        =  
        \int \Bigg( \prod_{i=1}^{L} \frac{d^Dk_i}{(2\pi)^D} \Bigg)
        \frac{\big(\mu^{4-D}\big)^L}{\mathcal{D}_1^{\nu_1} \cdots \mathcal{D}_N^{\nu_N}} \frac{-k_1^2k_2^4+D(k_1\cdot k_2)^2k_2^2}{(D-1)D(D+2)(D+4)}.
    \end{aligned}
\end{equation}
Hence, it is sufficient to contract the tensor integral 
with only one representative projector $P_{a}^{\mu_1\cdots\mu_6}$
for each coefficient $C_a$, where $P_{a}^{\mu_1\cdots\mu_6}$ 
admits the same index permutation symmetry as the respective term 
associated to the coefficient $C_a$.
Utilising the symmetries of the integrand consequently reduces the 
computational effort needed to extract the required scalar integrals.

In \texttt{FORM}, we implemented this procedure 
via the built-in functions \texttt{dd\_} and \texttt{distrib\_},
in order to identify the different tensors that build the terms invariant 
under the integrand's symmetry,
e.g.\ the two on the RHS of Eq.\ \eqref{Eq:TensorReduced-Rank6-TensorIntegral}
associated to $C_1$ and $C_4$, and to avoid generating 
duplicate terms by exploiting symmetries,
as suggested in Ref.\ \cite{Ruijl:2018poj}.

% -------------------------------------------------------------------------

% +++++++++++++++++++++++++++++++++++++++++++++++++++++++++++++++++++++++++
\section{Four-Loop Renormalisation}\label{Sec:4-Loop-Renormalisation}

Using the computational methods described 
in Sec.\ \ref{Sec:ComputationalSetup}, we calculate all relevant 
subrenormalised Green functions and present the results for the complete 4-loop renormalisation
of the chiral gauge theory discussed in the previous section.
As outlined, the theory is regularised within the framework of the BMHV scheme 
of DReg, leading to a spurious breaking of gauge and BRST invariance.
We construct the full 4-loop counterterm action, 
including the finite symmetry-restoring part, 
which renders the renormalised theory both finite and symmetric at this order.
All 4-loop contributions have been computed in Feynman gauge, i.e.\ $\xi=1$, 
and the anomaly cancellation condition, Eq.\ \eqref{Eq:AnomalyCancellationCondition}, 
has been imposed throughout.
For a complete list of counterterms at the 1-, 2- and 3-loop level, we refer the reader to Refs.\ \cite{Belusca-Maito:2021lnk,Stockinger:2023ndm}.

The counterterm action $S_{\mathrm{ct}}$ admits a decomposition into 
a divergent part, $S_{\mathrm{sct}}$,
and a finite, symmetry-restoring part, $S_{\mathrm{fct}}$.
The divergent piece itself can further be split 
into a symmetric part, $S_{\mathrm{sct,inv}}$,
and a BRST-breaking part, $S_{\mathrm{sct,break}}$.
Hence, one obtains
\begin{equation}\label{Eq:Full-CT-Action}
    \begin{aligned}
        S_{\mathrm{ct}} 
        = S_{\mathrm{sct}} + S_{\mathrm{fct}}
        = S_{\mathrm{sct,inv}} + S_{\mathrm{sct,break}} + S_{\mathrm{fct}}.
    \end{aligned}
\end{equation}
This counterterm action $S_{\mathrm{ct}}$, 
as well as its individual components 
$S_{\mathrm{sct,inv}}$, $S_{\mathrm{sct,break}}$ and $S_{\mathrm{fct}}$,
admits a perturbative expansion of the form
\begin{equation}\label{Eq:CT-Action-PerturbativeExpansion}
    \begin{aligned}
        S_{\mathrm{ct}} 
        = S^{1}_{\mathrm{ct}} + S^{2}_{\mathrm{ct}} + S^{3}_{\mathrm{ct}} 
        + S^{4}_{\mathrm{ct}} + \mathcal{O}(\text{5-loop}).
    \end{aligned}
\end{equation}

In Sec.\ \ref{Sec:4-Loop-Ssct}, we provide the 4-loop results for the complete singular counterterm action, 
derived from the power-counting divergent 
1PI Green functions listed below.
These counterterms ensure UV finiteness of the theory at the 4-loop order.

Next, in Sec.\ \ref{Sec:4-Loop-BRST-Breaking}, we discuss the  
4-loop breaking of BRST symmetry. 
For this purpose, we concisely illustrate our methodology
for symmetry restoration, based on quantum action principle of DReg,
as explained in Refs.\ 
\cite{Belusca-Maito:2020ala,Belusca-Maito:2021lnk,Stockinger:2023ndm,Kuhler:2024fak,Ebert:2024xpy}
(see Ref.\ \cite{Belusca-Maito:2023wah} for a review).
In particular, we extract the BRST-breaking contributions from
1PI Green functions with an insertion of a local $\Delta$-operator, 
and provide explicit results for all relevant Green functions.

Using these results, we ultimately derive 
the complete set of finite, symmetry-restoring 4-loop counterterms
in Sec.\ \ref{Sec:4-Loop-Sfct}.
Combined with the BRST-breaking part of the singular counterterm action,
these %finite 
counterterms restore the broken BRST symmetry,
such that the renormalised theory satisfies the Slavnov-Taylor identity 
at the given order.

As outlined in Sec.\ \ref{Sec:Setup}, 
we thoroughly tested our setup in various ways. 
Moreover, the UV-divergent BRST-breaking contributions serve as a strong 
consistency check of our results, as they are obtained
from both the standard 1PI Green functions provided in Sec.\ 
\ref{Sec:4-Loop-Ssct} and $\Delta$-inserted 1PI Green functions 
shown in Sec.\ \ref{Sec:4-Loop-BRST-Breaking}.
We find perfect agreement between both calculations.
Additionally, all counterterms are local polynomials in external momenta, 
as they have to be as a matter of principle, which constitutes a further 
non-trivial cross-check and provides confidence in our results.

\subsection{Four-Loop Singular Counterterm Action}\label{Sec:4-Loop-Ssct}

We begin by providing all power-counting divergent, standard 1PI Green functions
required to determine the singular counterterms.
In the Abelian chiral gauge theory considered here, see Sec.\ \ref{Sec:TheoryDefinition}, we obtain
\begin{align}
    \begin{split}\label{Eq:GF-BB}
        i \Gamma_{B_{\nu}(-p)B_{\mu}(p)} \big|_{\text{div}}^{4} &= 
        - \frac{i g^{8}}{(16 \pi^2)^4} \,
        \delta Z^{(4)}_{B}
        \Big(\overline{p}^{\mu} \overline{p}^{\nu} - \overline{p}^2 \overline{\eta}^{\mu\nu}\Big)\\
        &\phantom{-\,\,} + \frac{i g^{8}}{(16 \pi^2)^4} \Big(
        \delta \widehat{X}^{(4)}_{BB} \,
        \widehat{p}^2 \, \overline{\eta}^{\mu\nu}
        + \delta \overline{X}^{(4)}_{BB} \,
        \overline{p}^2 \, \overline{\eta}^{\mu\nu} \Big),
    \end{split}\\
    \begin{split}
        i \Gamma_{\psi_{j}(p)\overline{\psi}_{i}(-p)} \big|_{\text{div}}^{4} &= 
        - \frac{i g^{8}}{(16 \pi^2)^4} 
        \Big( 
        \delta Z^{(4)}_{\psi,ij}
        + \delta \overline{X}^{(4)}_{\overline{\psi}\psi,ij}
        \Big) \overline{\slashed{p}} \, \mathbb{P}_{\mathrm{R}},
    \end{split}\\
    \begin{split}
        i \Gamma_{\psi_j(p_2)\overline{\psi}_i(p_1)B_{\mu}(q)}\big|_{\text{div}}^{4} &= 
        \frac{i g^{9}}{(16 \pi^2)^4} \,
        \big(\mathcal{Y}_R\big)_{ik} \delta Z^{(4)}_{\psi,kj} \, 
        \overline{\gamma}^{\mu} \mathbb{P}_{\mathrm{R}},
    \end{split}\\
    \begin{split}
        i \Gamma_{B_{\mu}(q)B_{\nu}(p_1)B_{\rho}(p_2)} \big|_{\text{div}}^{4} &= 0,
    \end{split}\\
    \begin{split}\label{Eq:GF-BBBB}
        i \Gamma_{B_{\mu}(p_2)B_{\nu}(p_1)B_{\rho}(p_4)B_{\sigma}(p_3)} \big|_{\text{div}}^{4} &= 
        - \frac{i g^{10}}{(16 \pi^2)^4} \,
        \delta \overline{X}^{(4)}_{BBBB}
        \Big( \overline{\eta}^{\mu\nu} \, \overline{\eta}^{\rho\sigma} + \overline{\eta}^{\mu\rho} \, \overline{\eta}^{\nu\sigma} + \overline{\eta}^{\mu\sigma} \, \overline{\eta}^{\nu\rho} \Big),
    \end{split}
\end{align}
where $\Gamma$ is the effective quantum action and
all results are expressed in terms of counterterm coefficients.
In particular, for the BRST-invariant contributions their Laurent expansions in $\epsilon$ are given by
\begin{align}
    \begin{split}\label{Eq:CountertermCoeff-GaugeBoson-Inv}
        \delta Z^{(4)}_{B} &= 
          \mathcal{A}_{BB}^{4,\text{inv}} \, \frac{1}{\epsilon}
        + \mathcal{B}_{BB}^{4,\text{inv}} \, \frac{1}{\epsilon^2} 
        + \mathcal{C}_{BB}^{4,\text{inv}} \, \frac{1}{\epsilon^3},
    \end{split}\\
    \begin{split}\label{Eq:CountertermCoeff-Fermion-Inv}
        \delta Z^{(4)}_{\psi,ij} &= 
          \mathcal{A}_{\overline{\psi}\psi, ij}^{4,\text{inv}} \, \frac{1}{\epsilon}
        + \mathcal{B}_{\overline{\psi}\psi, ij}^{4,\text{inv}} \, \frac{1}{\epsilon^2}
        + \mathcal{C}_{\overline{\psi}\psi, ij}^{4,\text{inv}} \, \frac{1}{\epsilon^3}
        + \mathcal{D}_{\overline{\psi}\psi, ij}^{4,\text{inv}} \, \frac{1}{\epsilon^4},
    \end{split}
\end{align}
while for the BRST-breaking contributions we define
\begin{align}
    \begin{split}\label{Eq:CountertermCoeff-GaugeBoson-Break-Evanescent}
        \delta \widehat{X}^{(4)}_{BB} &=  
          \widehat{\mathcal{A}}_{BB}^{4,\text{break}} \, \frac{1}{\epsilon}
        + \widehat{\mathcal{B}}_{BB}^{4,\text{break}} \, \frac{1}{\epsilon^2} 
        + \widehat{\mathcal{C}}_{BB}^{4,\text{break}} \, \frac{1}{\epsilon^3}
        + \widehat{\mathcal{D}}_{BB}^{4,\text{break}} \, \frac{1}{\epsilon^4},
    \end{split}\\
    \begin{split}\label{Eq:CountertermCoeff-GaugeBoson-Break-4dim}
        \delta \overline{X}^{(4)}_{BB} &= 
          \overline{\mathcal{A}}_{BB}^{4,\text{break}} \, \frac{1}{\epsilon}
        + \overline{\mathcal{B}}_{BB}^{4,\text{break}} \, \frac{1}{\epsilon^2},
    \end{split}\\
    \begin{split}\label{Eq:CountertermCoeff-BBBB-Break}
        \delta \overline{X}^{(4)}_{BBBB} &=
          \overline{\mathcal{A}}_{BBBB}^{4,\text{break}} \, \frac{1}{\epsilon}
        + \overline{\mathcal{B}}_{BBBB}^{4,\text{break}} \, \frac{1}{\epsilon^2},
    \end{split}\\
    \begin{split}\label{Eq:CountertermCoeff-Fermion-Break}
        \delta \overline{X}^{(4)}_{\overline{\psi}\psi,ij} &= 
          \overline{\mathcal{A}}_{\overline{\psi}\psi, ij}^{4,\text{break}} \, \frac{1}{\epsilon}
        + \overline{\mathcal{B}}_{\overline{\psi}\psi, ij}^{4,\text{break}} \, \frac{1}{\epsilon^2}
        + \overline{\mathcal{C}}_{\overline{\psi}\psi, ij}^{4,\text{break}} \, \frac{1}{\epsilon^3}.
    \end{split}
\end{align}
We obtained analytical results for all of these pole coefficients in terms of the hypercharge matrix and Riemann zeta values.
The explicit expressions are somewhat lengthy and therefore provided in App.\ \ref{App:ExplicitResults}.

From these 1PI Green functions, we can now derive the complete singular counterterm action.
On the one side, the BRST-invariant part takes the form
\begin{equation}\label{Eq:Ssct_inv_4-Loop}
    \begin{aligned}
        S^{4}_{\mathrm{sct,inv}} = 
        \frac{g^8}{(16 \pi^2)^4} \Dintx 
        \bigg\{ 
        \delta Z^{(4)}_{B} \Big( -\frac{1}{4} \, \overline{F}^{\mu\nu} \, \overline{F}_{\mu\nu} \Big)
        + \delta Z^{(4)}_{\psi,kj} \overline{\psi}_i i \overline{\slashed{D}}_{\mathrm{R},ik} \psi_j
        \bigg\},
    \end{aligned}
\end{equation}
with right-handed covariant derivative
\begin{equation}
    \begin{aligned}
        \overline{D}^{\mu}_{\mathrm{R},ij} = \Big(
        \overline{\partial}^{\mu} \delta_{ij}
        + i g {\hypR}_{,ij} \overline{B}^{\mu}
        \Big) \projR.
    \end{aligned}
\end{equation}
On the other side, the BRST-breaking part of the singular counterterm action 
reads
\begin{equation}\label{Eq:Ssct_break_4-Loop}
    \begin{aligned}
        S^{4}_{\mathrm{sct,break}} = 
        \frac{g^8}{(16 \pi^2)^4} \Dintx 
        \bigg\{ 
        &\delta \widehat{X}^{(4)}_{BB} \, \frac{1}{2} \overline{B}_{\mu} \widehat{\partial}^2 \overline{B}^{\mu}
        + \delta \overline{X}^{(4)}_{BB} \, \frac{1}{2} \overline{B}_{\mu} \overline{\partial}^2 \overline{B}^{\mu}\\
        + \, &g^2 \delta \overline{X}^{(4)}_{BBBB} \, \frac{1}{8} \overline{B}_{\mu} \overline{B}^{\mu} \overline{B}_{\nu} \overline{B}^{\nu}
        + \delta \overline{X}^{(4)}_{\overline{\psi}\psi,ij} \Big( \overline{\psi}_i i \overline{\slashed{\partial}} \projR \psi_j \Big)
        \bigg\}.
    \end{aligned}
\end{equation}

It is straightforward to see that Eq.\ \eqref{Eq:Ssct_inv_4-Loop}
is BRST-invariant, given the structure of the field monomials.
Further, $S^{4}_{\mathrm{sct,inv}}$ 
retains the same structure as at lower orders, 
see Refs.\ \cite{Belusca-Maito:2021lnk,Stockinger:2023ndm}.
In fact, its principal structure is the same as in ordinary 
vector-like QED, differing only in the use
of a right-handed covariant derivative and modified coefficients.
This is because it consists only of symmetric field combinations
present in the tree-level Lagrangian and
corresponds to a multiplicative field and parameter
renormalisation of the gauge boson and the fermion fields, 
as well as the gauge coupling, such that the product 
$g B^{\mu}$ does not renormalise, just as in ordinary QED.

The BRST-breaking part of the counterterm action, given
in Eq.\ \eqref{Eq:Ssct_break_4-Loop}, does also not introduce any new 
field monomials compared to the 3-loop case, cf.\ Ref.\ \cite{Stockinger:2023ndm}.
This contrasts with the transitions from 1 to 2 loops and from 2 to 3 loops, 
where new field monomials emerged at the higher loop order.
Hence, the principal structure of the 4-loop BRST-breaking singular counterterm
action remains the same as at the 3-loop order, with the only difference
being the presence of higher order $\epsilon$-poles in the counterterm coefficients.
The absence of new field monomials is due to constraints imposed by power-counting and renormalisability,
as well as the absence of evanescent gauge and Yukawa interactions, 
see also Ref.\ \cite{Ebert:2024xpy}.

In analogy to the discussion in Ref.\ \cite{Stockinger:2023ndm},
we choose to attribute the BRST breaking entirely to the bilinear 
fermion terms, see Eq.\ \eqref{Eq:Ssct_break_4-Loop}.
However, in the present context, this is just a choice and we also could have chosen the 
fermion-gauge boson vertex correction to carry the BRST breaking
(or distribute it between both) using a slight rearrangement.
In particular, we could rewrite these terms as
\begin{equation}
    \begin{aligned}
        \frac{g^8}{(16 \pi^2)^4} \Dintx 
        \bigg\{
        \Big( 
        \delta Z^{(4)}_{\psi,kj} 
        + \delta \overline{X}^{(4)}_{\overline{\psi}\psi,kj} \Big) 
        \overline{\psi}_i i \overline{\slashed{D}}_{\mathrm{R},ik} \psi_j
        + g {\hypR}_{,ik}
        \delta \overline{X}^{(4)}_{\overline{\psi}\psi,kj} 
        \overline{\psi}_i \overline{\slashed{B}} \projR \psi_j
        \bigg\},
    \end{aligned}
\end{equation}
which leads to the same counterterm action, $S^{4}_{\mathrm{sct}}$,
but with the BRST breaking fully attributed to the gauge interaction.
We emphasise that this freedom depends on the specific model under consideration, as will be discussed in more detail at the end of Sec.~\ref{Sec:4-Loop-Sfct}.

\subsection{Four-Loop BRST Breaking}\label{Sec:4-Loop-BRST-Breaking}

Ultimately, we require the renormalised theory in 4 dimensions to satisfy the 
Slavnov-Taylor identity, i.e.\
\begin{equation}\label{Eq:UltimateSymmetryRequirement}
    \begin{aligned}
        \mathop{\text{LIM}}_{D \, \to \, 4} \, (\mathcal{S}_D(\Gamma_\mathrm{DRen})) = 0,
    \end{aligned}
\end{equation}
where $\mathcal{S}_D$ is the $D$-dimensional Slavnov-Taylor operator, 
$\Gamma_\mathrm{DRen}$ denotes the renormalised effective quantum action in 
$D$ dimensions and $\mathop{\text{LIM}}_{D \, \to \, 4}$ includes 
the limit $D \to 4$ and dropping finite evanescent terms, i.e.\ evanescent Lorentz structures whose coefficients contain no poles in $\epsilon$.
The symmetry-restoring counterterms must be adjusted such that Eq.\ \eqref{Eq:UltimateSymmetryRequirement} becomes valid and thus the breaking induced by the BMHV regularisation is cancelled. 
As the crucial preliminary step, the breaking of Eq.\ \eqref{Eq:UltimateSymmetryRequirement} on the regularised level has to be evaluated.
Following Refs.\
\cite{Belusca-Maito:2020ala,Belusca-Maito:2021lnk,Belusca-Maito:2023wah,Stockinger:2023ndm,Kuhler:2024fak,Ebert:2024xpy},
an efficient way to obtain this breaking
is to
utilise the quantum action principle of DReg, see Refs.\
\cite{Breitenlohner:1975hg,Breitenlohner:1976te,Breitenlohner:1977hr} and
Ref.\ \cite{Belusca-Maito:2023wah} for a review, 
and rewrite the symmetry-breaking as an insertion of
a local composite operator as
\begin{equation}\label{Eq:QAPofDReg}
    \begin{aligned}
        \mathcal{S}_D(\Gamma_\mathrm{DRen})=\Delta\cdot\Gamma_\mathrm{DRen}.
        %\mathcal{S}_D(\Gamma)=\Delta\cdot\Gamma.
    \end{aligned}
\end{equation}
The $\Delta$-operator on the RHS of Eq.\ (\ref{Eq:QAPofDReg}) is defined by
\begin{equation}\label{Eq:DefDeltaBreaking}
    \begin{aligned}
        \Delta = \widehat{\Delta} + \Delta_{\mathrm{ct}} = \mathcal{S}_D(S_{0} + S_{\mathrm{ct}}).
    \end{aligned}
\end{equation}
For practical perturbative calculations, we plug Eq.\ \eqref{Eq:QAPofDReg}
into Eq.\ \eqref{Eq:UltimateSymmetryRequirement} and obtain the 
perturbative requirement
\begin{equation}\label{Eq:PerturbativeRequirementAndStartingPoint}
    \begin{aligned}
        \mathop{\text{LIM}}_{D \, \to \, 4} \, \bigg(\widehat{\Delta}\cdot\Gamma_\mathrm{DRen}^L+\sum_{k=1}^{L-1}\Delta^k_\mathrm{ct}\cdot\Gamma^{L-k}_\mathrm{DRen}+\Delta^L_\mathrm{ct}\bigg)=0, \hspace{0.75cm} \forall \, L \geq 1,
    \end{aligned}
\end{equation}
where $L$ is the loop order of the respective quantities
and the local composite operator $\Delta$ admits the following 
perturbative expansion:
\begin{equation}\label{Eq:DeltaOperator-PerturbativeExpansion}
    \begin{aligned}
        \Delta = 
        \widehat{\Delta} + \Delta^1_\mathrm{ct}
        + \Delta^2_\mathrm{ct} + \Delta^3_\mathrm{ct}
        + \Delta^4_\mathrm{ct} + \mathcal{O}(\text{5-loop}).
    \end{aligned}
\end{equation}
While $\widehat{\Delta}$ represents the breaking of BRST symmetry 
at tree-level, the operator $\Delta^L_\mathrm{ct}$ reflects 
the breaking at loop order $L$.
Hence, using Eq.\ \eqref{Eq:PerturbativeRequirementAndStartingPoint}
as a starting point, we can compute the 4-loop BRST breaking 
$\Delta^4_\mathrm{ct}$ from\footnote{
Here, we dropped the subscript ``$\mathrm{subren}$''. 
Note that $\Gamma_\mathrm{subren}$ denotes the subrenormalised 
effective quantum action in $D$ dimensions which, 
in contrast to $\Gamma_\mathrm{DRen}$, 
does not include the genuine $4$-loop counterterms. 
}
\begin{equation}\label{Eq:4-loop-BRST-Breaking}
    \begin{aligned}
        \big( \Delta \cdot \Gamma \big)^{4} = 
        \widehat{\Delta} \cdot \Gamma^{4}
        + \Delta^1_\mathrm{ct} \cdot \Gamma^{3}
        + \Delta^2_\mathrm{ct} \cdot \Gamma^{2}
        + \Delta^3_\mathrm{ct} \cdot \Gamma^{1} 
        = - \Delta^4_\mathrm{ct} + \mathcal{O}(\hat{.}),
    \end{aligned}
\end{equation}
where $\mathcal{O}(\hat{.})$ indicates finite evanescent terms that
ultimately drop 
out in the limit $\mathop{\text{LIM}}_{D\,\to\,4}$ in 
Eq.\ \eqref{Eq:UltimateSymmetryRequirement}.
Notably, to obtain the complete result for the BRST breaking,
including both divergent and finite contributions,
we can restrict ourselves to power-counting divergent,
subrenormalised 1PI Green functions with an insertion of the
$\Delta$-operator.
The reason for this is that the lowest-order part of the $\Delta$-operator, 
i.e.\ $\widehat{\Delta}$ arising from the tree-level breaking 
of BRST invariance, is purely evanescent.
This represents an immense simplification and is one of the
main reasons for the efficiency of our method.

For the Abelian theory considered in this work, see Sec.\ \ref{Sec:TheoryDefinition},
we obtain these $\Delta$-inserted 1PI Green functions at the 4-loop level as
\begin{align}
    \begin{split}\label{Eq:GF-cB}
        i \Delta \cdot \Gamma \big|_{B_{\mu}(-p) c(p)}^{4} &= \frac{g^{8}}{(16 \pi^2)^4} 
        \bigg[ 
        \delta \widehat{X}^{(4)}_{BB} \, 
        \widehat{p}^{2} \, \overline{p}^{\mu}
        + \Big( 
        \delta \overline{X}^{(4)}_{BB}  
        + \mathcal{F}_{BB}^{4,\text{break}}
        \Big) 
        \overline{p}^{2} \, \overline{p}^{\mu}
        \bigg],
    \end{split}\\
    \begin{split}
        i \Delta \cdot \Gamma \big|_{\psi_j(p_2) \overline{\psi}_i(p_1) c(q)}^{4}
        &= \frac{g^{9}}{(16 \pi^2)^4} \big(\mathcal{Y}_R\big)_{ik} 
        \Big(
        \delta \overline{X}^{(4)}_{\overline{\psi}\psi,kj}
        + \mathcal{F}_{\psi\overline{\psi}, kj}^{4,\text{break}} \Big) 
        \big( \overline{\slashed{p}}_1 + \overline{\slashed{p}}_2 \big) \mathbb{P}_{\mathrm{R}},
    \end{split}\\
    \begin{split}
        i \Delta \cdot \Gamma \big|_{B_{\nu}(p_2) B_{\mu}(p_1) c(q)}^{4} &= 0,
    \end{split}\\
    \begin{split}\label{Eq:GF-cBBB}
        i \Delta \cdot \Gamma \big|_{B_{\rho}(p_3) B_{\nu}(p_2) B_{\mu}(p_1) c(q)}^{4} &= 
        \frac{g^{10}}{(16 \pi^2)^4} 
        \Big( 
        \delta \overline{X}^{(4)}_{BBBB} 
        + \mathcal{F}_{BBBB}^{4,\text{break}} 
        \Big) \\
        &\times \big(\overline{p}_1 + \overline{p}_2 + \overline{p}_3\big)_{\sigma} \, \big( \overline{\eta}^{\mu\nu} \, \overline{\eta}^{\rho\sigma} + \overline{\eta}^{\mu\rho} \, \overline{\eta}^{\nu\sigma} + \overline{\eta}^{\mu\sigma} \, \overline{\eta}^{\nu\rho} \big),
    \end{split}
\end{align}
with divergent counterterm coefficients as defined in Eqs.\ 
\eqref{Eq:CountertermCoeff-GaugeBoson-Break-Evanescent} to
\eqref{Eq:CountertermCoeff-Fermion-Break}.
Our explicit analytical results for the finite 
counterterm coefficients are provided in App.\ \ref{App:ExplicitResults}.

Note that we do not consider the two power-counting divergent 1PI Green functions
\begin{align}
    \begin{split}\label{Eq:GF-cFBF}
        i \Delta \cdot \Gamma \big|_{\psi_{j}(p_3) \overline{\psi}_i(p_2) B_{\mu}(p_1) c(q)}^{4} &\equiv 0,
    \end{split}\\
    \begin{split}\label{Eq:GF-cBBBB}
        i \Delta \cdot \Gamma \big|_{B_{\sigma}(p_4) B_{\rho}(p_3) B_{\nu}(p_2) B_{\mu}(p_1) c(q)}^{4} &\equiv 0,
    \end{split}
\end{align}
as they vanish identically.
In the Abelian theory studied here, this results from a cancellation
of the leading power-counting term in the 
respective integrands.
For the latter Green function, see Eq.\ \eqref{Eq:GF-cBBBB}, 
this also follows from renormalisability arguments, 
as discussed in Refs.\ \cite{Stockinger:2023ndm,Ebert:2024xpy}. 

Indeed, these Green functions fully capture the breaking of BRST symmetry
at the 4-loop level, both divergent and finite, 
as reflected by the breaking of the 
Slavnov-Taylor identity according to Eq.\ \eqref{Eq:QAPofDReg}.
Consequently, the full 4-loop breaking of BRST symmetry can be derived as
\begin{equation}\label{Eq:Full-BRST-Breaking-4-Loop}
    \begin{aligned}
        \big( \Delta \cdot \Gamma \big)^{4}
        = \frac{g^8}{(16 \pi^2)^4} &\Dintx \, \bigg\{
        \delta \widehat{X}^{(4)}_{BB} \, c \widehat{\partial}^2 \overline{\partial}^{\mu} \overline{B}_{\mu}
        + \Big[ \delta \overline{X}^{(4)}_{BB} + \mathcal{F}_{BB}^{4,\text{break}} \Big]
        c \overline{\partial}^2 \overline{\partial}^{\mu} \overline{B}_{\mu}\\
        %&\quad\,\,
        &\qquad\,\,
        + g^2 \Big[ \delta \overline{X}^{(4)}_{BBBB} + \mathcal{F}_{BBBB}^{4,\text{break}} \Big] 
        \frac{1}{2} c \, \overline{\partial}_{\mu} \big(\overline{B}^{\mu}\overline{B}_{\nu}\overline{B}^{\nu}\big)\\
        %&\quad\,\,
        &\qquad\,\,
        + g {\hypR}_{,ik}
        \Big[ \delta \overline{X}^{(4)}_{\overline{\psi}\psi,kj} + \mathcal{F}_{\overline{\psi}\psi,kj}^{4,\text{break}} \Big]
        c \, \overline{\partial}_{\mu} \big( \overline{\psi}_i \overline{\gamma}^{\mu} \projR \psi_j \big)
        \bigg\}
         + \mathcal{O}(\hat{.}),
    \end{aligned}
\end{equation}
omitting finite evanescent terms.
From this result, we obtain 
$\Delta^4_\mathrm{ct}$ via Eq.\ \eqref{Eq:4-loop-BRST-Breaking}.

Using Eq.\ \eqref{Eq:DefDeltaBreaking},
we can first verify that the divergent contributions of
$\Delta^4_\mathrm{ct}$ indeed correctly reproduce the BRST-breaking part
of the singular counterterm action, given in Eq.\
\eqref{Eq:Ssct_break_4-Loop}, serving as the aforementioned 
consistency check.
Second, and more important, we can finally derive the finite,
symmetry-restoring counterterms from the finite contributions 
of $\Delta^4_\mathrm{ct}$, which are discussed in the following subsection.

\subsection{Four-Loop Finite Symmetry-Restoring Counterterms}\label{Sec:4-Loop-Sfct}

From the finite contributions to the BRST breaking in Eq.\
\eqref{Eq:Full-BRST-Breaking-4-Loop},
we ultimately derive the finite, symmetry-restoring counterterms
at the 4-loop level.
Using Eqs.\ \eqref{Eq:DefDeltaBreaking} and \eqref{Eq:4-loop-BRST-Breaking}, we obtain 
\begin{equation}\label{Eq:Sfct-4-Loop}
    \begin{aligned}
        S^{4}_{\mathrm{fct}} = 
        \frac{g^8}{(16 \pi^2)^4} \intx 
        \bigg\{ 
        &\mathcal{F}_{BB}^{4,\text{break}} \, \frac{1}{2} \overline{B}_{\mu} \overline{\partial}^2 \overline{B}^{\mu} + g^2 \mathcal{F}_{BBBB}^{4,\text{break}} \, \frac{1}{8} \overline{B}_{\mu} \overline{B}^{\mu} \overline{B}_{\nu} \overline{B}^{\nu}\\
        + \, &
        \mathcal{F}_{\overline{\psi}\psi,ij}^{4,\text{break}} \, \overline{\psi}_i i \overline{\slashed{\partial}} \projR \psi_j
        \bigg\},
    \end{aligned}
\end{equation}
where our explicit, analytical results for the finite counterterm coefficients in terms of the hypercharge matrix and Riemann zeta values are provided in App.\ \ref{App:ExplicitResults}.
This is the main result of this paper.

Together with $S^{4}_{\mathrm{sct,break}}$ in Eq.\ \eqref{Eq:Ssct_break_4-Loop},
these counterterms restore the broken BRST symmetry at the 4-loop level, 
guaranteeing that the Slavnov-Taylor identity, 
see Eq.\ \eqref{Eq:UltimateSymmetryRequirement},
is satisfied at that order in perturbation theory.
Analogous to the situation at lower loop orders $L\!\leq\!3$, there
are no new field monomials in $S_{\mathrm{fct}}$ emerging at the 4-loop level,
as expected and discussed in Ref.\ \cite{Stockinger:2023ndm}.
Hence, the principal structure of the finite, symmetry-restoring counterterm 
action remains unchanged at the 4-loop order and, restricted by power-counting, 
is expected to be retained also at higher orders.

As at lower loop orders, the symmetry-restoring counterterm action
is not unique, since only its BRST variation in 
$\mathop{\text{LIM}}_{D \, \to \, 4}$ is fixed.
Our choice at 4 loops is consistent with that employed at 
lower orders, see Ref.\ \cite{Stockinger:2023ndm}.
First, we restrict to purely $4$-dimensional operators.
Second, we have chosen not to use a field monomial of the form 
$\overline{\psi}B\psi$,
thereby assigning the entire 
fermionic breaking to the bilinear fermion term.
This is possible due to the freedom to add any symmetric, 
finite counterterm to the action.
In particular, we may add the finite and symmetric counterterm
\begin{equation}
    \begin{aligned}
        - \frac{g^8}{(16 \pi^2)^4} \intx \, 
        \mathcal{F}_{\overline{\psi}\psi,kj}^{4,\text{break}} \,
        \overline{\psi}_i i \overline{\slashed{D}}_{\mathrm{R},ik} \psi_j
    \end{aligned}
\end{equation}
to attribute the breaking entirely to the fermion-gauge boson interaction
(or add a different finite, symmetric counterterm to distribute it between both).

It is important to note that we cannot always attribute 
the complete fermionic breaking to the fermion bilinear terms alone.
In Ref.\ \cite{Ebert:2024xpy} this was not possible 
due to non-vanishing contributions from
the 1PI Green function for $c\overline{\psi}B\psi$, see Eq.\ \eqref{Eq:GF-cFBF},
arising from evanescent gauge interactions and Yukawa contributions,
which are not present in the theory studied here.
Therefore, in Ref.\ \cite{Ebert:2024xpy}, the fermionic breaking was assigned to the fermion-gauge boson interaction,  
leaving out fermion bilinear terms,
in order to obtain a minimal expression for $S_{\mathrm{fct}}$.

\subsection{Application to the Abelian Fermionic Sector of the SM}\label{Sec:4-Loop-Abelian-SM}

The Abelian fermionic sector of the SM can be mapped to the notation used in this paper.
The main idea is that the left-handed 
physical fermions of the SM ${\psi_{L}}_i$ can be rewritten via equivalent 
right-handed, charge-conjugated spinor fields $({\psi_L}_i)^C$.
This amounts to working with ``Option 2b'' for
the treatment of fermions in $D$ dimensions, as introduced and explained
in Ref.\ \cite{Ebert:2024xpy}, Sec.\ 2.1.
In this way, the fermion multiplet $\psi_R$ consists of all the
physical fermions of the SM, which are
$\nu_I\in\{\nu_e,\nu_{\mu},\nu_{\tau}\}$,\footnote{
In this context, we work with sterile right-handed neutrinos $\nu_{R_I}^{\text{st}}$ 
and physical left-handed neutrinos $(\nu_{L_I})^C$, implemented via charge-conjugation as described above.
} $e_I\in\{e,\mu,\tau\}$, 
$u^i_I\in\{u^i,c^i,t^i\}$, $d^i_I\in\{d^i,s^i,b^i\}$, with generation index $I$ and colour index $i$.
In this framework, the hypercharges of the SM fermions are given by
the hypercharge matrix 
\begin{equation}\label{Eq:ASM-Hypercharges}
    \begin{aligned}
        \hypR^{\text{SM}} =
        \begin{pmatrix}
            0 & 0 & 0 & 0 &  &  &  & \\
            0 & - \mathbb{1}_{3\times3} & 0 & 0 &  &  &  & \\
            0 & 0 & \frac{2}{3} \mathbb{1}_{9\times9} & 0 &  &  &  & \\
            0 & 0 & 0 & -\frac{1}{3} \mathbb{1}_{9\times9} &  &  &  & \\
             &  &  &  & \frac{1}{2} \mathbb{1}_{3\times3} & 0 & 0 & 0\\
             &  &  &  & 0 & \frac{1}{2} \mathbb{1}_{3\times3} & 0 & 0\\
             &  &  &  & 0 & 0 & -\frac{1}{6} \mathbb{1}_{9\times9} & 0\\
             &  &  &  & 0 & 0 & 0 & -\frac{1}{6} \mathbb{1}_{9\times9}
        \end{pmatrix},
    \end{aligned}
\end{equation}
where the upper left block corresponds to the hypercharges of the 
right-handed fermions and the lower right block to the 
hypercharges of the charge-conjugated left-handed fermions.
Clearly, this matrix satisfies the anomaly cancellation condition
Eq.\ \eqref{Eq:AnomalyCancellationCondition}.
Consistent with the SM hypercharges, it also fulfills the gauge-gravitational 
anomaly cancellation condition, $\mathrm{Tr}(\hypR)=0$,
and yields $\mathrm{Tr}(\hypR^2)=10$. 
The latter reflects the fact that all three generations are combined 
into a single multiplet, and is associated with the normalisation factor 
of $\sqrt{5/3}$ for the hypercharge gauge coupling in the context 
of Grand Unified Theories.

Using the explicit Standard Model hypercharges $\hypR^{\text{SM}}$, 
see Eq.\ \eqref{Eq:ASM-Hypercharges}, and plugging them into
the finite counterterm coefficients in Eqs.\ \eqref{Eq:Finite-BB-Break}, 
\eqref{Eq:Finite-BBBB-Break} and \eqref{Eq:Finite-FF-Break},
we obtain
\begin{equation}\label{Eq:Sfct-4-Loop-SM}
    \begin{aligned}
        S^{4}_{\mathrm{fct}}\Big|_{\hypR=\hypR^{\text{SM}}} &= 
        \frac{g^8}{(16 \pi^2)^4} \intx 
        \bigg\{ 
        \mathcal{F}_{BB}^{4,\text{SM}} \, \frac{1}{2} \overline{B}_{\mu} \overline{\partial}^2 \overline{B}^{\mu} + g^2 \mathcal{F}_{BBBB}^{4,\text{SM}} \, \frac{1}{8} \overline{B}_{\mu} \overline{B}^{\mu} \overline{B}_{\nu} \overline{B}^{\nu}\\
        &\qquad\qquad + \mathcal{F}_{\overline{e}e}^{4,\text{SM}} \, \overline{e_{R_I}} i \overline{\slashed{\partial}} e_{R_I} 
        + \mathcal{F}_{\overline{u}u}^{4,\text{SM}} \, \overline{u^i_{R_I}} i \overline{\slashed{\partial}} u^i_{R_I}
        + \mathcal{F}_{\overline{d}d}^{4,\text{SM}} \, \overline{d^i_{R_I}} i \overline{\slashed{\partial}} d^i_{R_I}\\
        &\qquad\qquad + \mathcal{F}_{\overline{L}L}^{4,\text{SM}} \, \overline{\big(L_{L_I}\big)^C} i \overline{\slashed{\partial}} \big(L_{L_I}\big)^C
        + \mathcal{F}_{\overline{Q}Q}^{4,\text{SM}} \, \overline{\big(Q^i_{L_I}\big)^C} i \overline{\slashed{\partial}} \big(Q^i_{L_I}\big)^C
        \bigg\},
    \end{aligned}
\end{equation}
for the 4-loop finite, symmetry-restoring counterterms, with explicit coefficients given by
{\small
\begin{alignat*}{5}
    %\begin{split}
        \mathcal{F}_{BB}^{4,\text{SM}} &= \frac{105574465087}{604661760} &&+ \frac{2665684621}{6998400} \zeta(3) &&- \frac{499349}{174960} \zeta(4) &&- \frac{99009133}{209952} \zeta(5)
        &&\approx 140.38,\\
    %\end{split}\\
    %\begin{split}
        \mathcal{F}_{BBBB}^{4,\text{SM}} &= \frac{1740074889071}{27209779200} &&- \frac{3393600941}{7873200} \zeta(3) &&+ \frac{151740709}{3149280} \zeta(4) &&- \frac{50011855}{944784} \zeta(5)
        &&\approx -456.91,\\
    %\end{split}\\
    %\begin{split}
        \mathcal{F}_{\overline{e}e}^{4,\text{SM}} &= \frac{27665962754029}{75582720000} &&+ \frac{3480124631}{17496000} \zeta(3) &&+ \frac{211030883}{583200} \zeta(4) &&- \frac{20725669}{19440} \zeta(5)
        &&\approx -108.73,\\
    %\end{split}\\
    %\begin{split}
        \mathcal{F}_{\overline{u}u}^{4,\text{SM}} &= \frac{39860981881231}{680244480000} &&+ \frac{22394315941}{88573500} \zeta(3) &&+ \frac{66753553}{1312200} \zeta(4) &&- \frac{231672547}{787320} \zeta(5)
        &&\approx 112.46,\\
    %\end{split}\\
    %\begin{split}
        \mathcal{F}_{\overline{d}d}^{4,\text{SM}} &= \frac{2388867184943}{340122240000} &&- \frac{67097069}{1417176000}\zeta(3) &&+ \frac{291946387}{5248800} \zeta(4) &&- \frac{93959699}{1574640} \zeta(5)
        &&\approx 5.29,\\
    %\end{split}\\
    %\begin{split}
        \mathcal{F}_{\overline{L}L}^{4,\text{SM}} &= \frac{62763071106811}{4837294080000} &&+ \frac{144119506409}{1119744000} \zeta(3) &&+ \frac{619731137}{37324800} \zeta(4) &&- \frac{34588759}{311040} \zeta(5)
        &&\approx 70.35,\\
    %\end{split}\\
    %\begin{split}
        \mathcal{F}_{\overline{Q}Q}^{4,\text{SM}} &= \frac{72769737196819}{43535646720000} &&- \frac{437815601831}{90699264000} \zeta(3) &&+ \frac{7110059113}{335923200} \zeta(4) &&- \frac{476640839}{25194240} \zeta(5)
        &&\approx -0.84.
    %\end{split}
\end{alignat*}
}

% -------------------------------------------------------------------------

% +++++++++++++++++++++++++++++++++++++++++++++++++++++++++++++++++++++++++
\section{Conclusions}\label{Sec:Conclusion}

In this work, we successfully performed the first 4-loop renormalisation of an Abelian
chiral gauge theory within the BMHV scheme of DReg,
providing a fully self-consistent and systematic treatment of $\gamma_5$.
We computed the complete counterterm action, including both singular and finite, symmetry-restoring counterterms.
In particular, we highlight the finite, symmetry-restoring counterterm
action in Eq.\ \eqref{Eq:Sfct-4-Loop} as a key result of our work.
The feasibility of this calculation relied heavily on our efficient methodology and optimised computational framework.

In contrast to previous studies at lower loop orders $(L\leq3)$, 
no new field monomials 
emerge in the BRST-breaking contribution to the singular 
counterterm action at 4 loops.
The same holds for the finite, symmetry-restoring counterterms, 
where this outcome was in fact anticipated.
The absence of new field monomials is due to the non-existence 
of evanescent gauge and Yukawa interactions,
as well as constraints from power-counting and renormalisability.
The primary difference
at the 4-loop order is the increased complexity of the 
counterterm coefficients.
Thus, we have demonstrated that the complete counterterm action can be expressed 
in a relatively compact form, even at the 4-loop level, amenable to a straight-forward computer implementation.

In order to perform this calculation, we developed an efficient, automated computational setup.
We implemented various optimised algorithms in a \texttt{FORM} program, including
an efficient implementation of the BMHV algebra and a dedicated
procedure for handling $\gamma$-matrices.
Specifically, we derived and implemented a factorisation of $\gamma$-traces,
decomposing every trace into an evanescent and a 4-dimensional part.
This allows us to use well-established and highly efficient $\gamma$-trace
algorithms, even within the framework of the BMHV scheme.
Additionally, we implemented an efficient algorithm for the reduction 
of vacuum bubble tensor integrals employing the orbit partition approach.
Furthermore, we have systematically applied a concise
procedure for symmetry restoration based on the regularised quantum action principle of DReg, as established in earlier work.
Overall, this project required the calculation of various 4-loop
1PI Green functions, some of which involved billions of terms 
per Feynman diagram in intermediate steps.
The successful execution of these computations
not only demonstrates the practical feasibility of 
4-loop calculations in the BMHV scheme but also 
serves as a validation of our computational framework.

The successful application of the BMHV scheme at the 4-loop order demonstrated in this work suggests the suitability of this framework for automated, self-consistent, multi-loop computations in the Standard Model and beyond.
This sets the computational stage for more precise calculations of 
electroweak observables, addressing the increasingly stringent 
precision requirements of modern experiments.

% -------------------------------------------------------------------------

%%%%% Acknowledgments %%%%%
\section*{Acknowledgments}
D.S.\ and M.W.\ acknowledge financial support by the German Science 
Foundation DFG, grant STO 876/8-1,2.
We would like to thank
Amon Ilakovac and Paul K\"uhler for their insightful ideas 
and stimulating discussions. 
Additionally, we are grateful to Ben Ruijl for valuable discussions on the
tensor reduction and to York Schroeder for his help with the 4-loop
tadpole master integrals, in particular for providing their numerical solutions to perform consistency checks.
Furthermore, we thank Joshua Davies, Takahiro Ueda and Jos Vermaseren
for discussions and advice regarding \texttt{FORM} internal parameter setups.
We extend our gratitude to the Centre for 
Information Services and High Performance Computing $[$Zentrum für Informationsdienste und 
Hochleistungsrechnen (ZIH)$]$ TU Dresden for providing its facilities for high-performance calculations,
and particularly thank Ulf Markwardt for his support.
Finally, M.W.\ gratefully acknowledges funding from the Graduate Academy of TU Dresden 
for a three-month research stay at the University of Regensburg, during 
which a substantial part of this work was carried out.

All Feynman diagrams in this paper have been illustrated using \texttt{JaxoDraw} \cite{Binosi:2008ig}.

% +++++++++++++++++++++++++++++++++++++++++++++++++++++++++++++++++++++++++
\begin{appendices}
\addappheadtotoc
\clearpage

\section{Derivation of the $\gamma$-Trace Factorisation in $D$ Dimensions}\label{App:TraceFactorisation}

In this Appendix, we discuss the proof of the factorisation
of $\gamma$-traces in $D$ dimensions within the BMHV scheme.
In order to see that Eq.\ \eqref{Eq:TraceFactorisationTheorem} holds, 
we consider all three occurring traces individually.
Further, we focus on the case $\mathbb{\Lambda}=\mathbb{1}$
and provide arguments for the generalisation to 
$\mathbb{\Lambda}\in\{\mathbb{1},\gamma_5,\projL,\projR\}$ below.

We begin with the $4$-dimensional traces 
$\mathrm{Tr}(\overline{\gamma}^{\mu_1}\ldots\overline{\gamma}^{\mu_n}\mathbb{1})$
for which we obtain
\begin{equation}\label{Eq:GammaTrace-4-Dim}
    \begin{aligned}
    \mathrm{Tr}(\overline{\gamma}^{\mu_1}\ldots\overline{\gamma}^{\mu_n})
    &= \mathrm{Tr}(\overline{\gamma}^{\mu_2}\ldots\overline{\gamma}^{\mu_n}\overline{\gamma}^{\mu_1})
    = \mathrm{Tr}\Big(\frac{1}{2}\{\overline{\gamma}^{\mu_1},\overline{\gamma}^{\mu_2}\ldots\overline{\gamma}^{\mu_n}\}\Big)\\
    &= \sum_{k=2}^{n} (-1)^k \, \overline{\eta}^{\mu_1\mu_k} \, \mathrm{Tr}(\overline{\gamma}^{\mu_2}\ldots\overline{\gamma}^{\mu_{k-1}}\overline{\gamma}^{\mu_{k+1}}\ldots\overline{\gamma}^{\mu_{n}})\\
    &= \sum_{k=1}^{(n-1)!!} (-1)^{\sigma_k(1,\ldots,n)} \, \overline{\eta}^{\mu_{\sigma_k(1)}\mu_{\sigma_k(2)}}\ldots\overline{\eta}^{\mu_{\sigma_k(n-1)}\mu_{\sigma_k(n)}} \, \mathrm{Tr}(\mathbb{1})\\
    &= 4 \sum_{k=1}^{(n-1)!!} (-1)^{\sigma_k(1,\ldots,n)} \, \overline{\eta}^{\mu_{\sigma_k(1)}\mu_{\sigma_k(2)}}\ldots\overline{\eta}^{\mu_{\sigma_k(n-1)}\mu_{\sigma_k(n)}},
    \end{aligned}
\end{equation}
which is a well-known formula.
Here, we used cyclicity of the trace in the first line,
evaluated the anticommutator in the second line (see Eq.\ \eqref{Eq:BMHV-Algebra-1}),
went to the third line via recursion and 
used Eq.\ \eqref{Eq:GammaTraceDefinitions} in the last step.
Further, we have $(n-1)!!=n!/(2^{n/2}(n/2)!)$, as well as
\begin{equation}
    \begin{aligned}
        (-1)^{\sigma_k(1,\ldots,n)} = 
        \begin{cases}
            1, &\text{for even permutations of $\{1,\ldots,n\}$},\\
            -1, &\text{for odd permutations of $\{1,\ldots,n\}$}.\\
        \end{cases}
    \end{aligned}
\end{equation}
Similarly, for the evanescent $\gamma$-traces, we find
\begin{equation}\label{Eq:GammaTrace-Evanescent}
    \begin{aligned}
    \mathrm{Tr}(\widehat{\gamma}^{\nu_1}\ldots\widehat{\gamma}^{\nu_m})
    &= \sum_{k=2}^{m} (-1)^k \, \widehat{\eta}^{\nu_1\nu_k} \, \mathrm{Tr}(\widehat{\gamma}^{\nu_2}\ldots\widehat{\gamma}^{\nu_{k-1}}\widehat{\gamma}^{\nu_{k+1}}\ldots\widehat{\gamma}^{\nu_{m}})\\
    %&= \sum_{k=1}^{(m-1)!!} (-1)^{\sigma_k(1,\ldots,m)} \, \widehat{\eta}^{\nu_{\sigma_k(1)}\nu_{\sigma_k(2)}}\ldots\widehat{\eta}^{\nu_{\sigma_k(m-1)}\nu_{\sigma_k(m)}} \, \mathrm{Tr}(\mathbb{1})\\
    &= 4 \sum_{k=1}^{(m-1)!!} (-1)^{\sigma_k(1,\ldots,m)} \, \widehat{\eta}^{\nu_{\sigma_k(1)}\nu_{\sigma_k(2)}}\ldots\widehat{\eta}^{\nu_{\sigma_k(m-1)}\nu_{\sigma_k(m)}}.
    \end{aligned}
\end{equation}
Finally, the LHS of Eq.\ \eqref{Eq:TraceFactorisationTheorem}
evaluates to
\begin{align}\label{Eq:GammaTrace-General}
    &\mathrm{Tr}(\widehat{\gamma}^{\nu_1}\ldots\widehat{\gamma}^{\nu_m}\overline{\gamma}^{\mu_1}\ldots\overline{\gamma}^{\mu_n}) \nonumber\\
    &= (-1)^{n+m} \, \mathrm{Tr}(\widehat{\gamma}^{\nu_2}\ldots\widehat{\gamma}^{\nu_m}\widehat{\gamma}^{\nu_1}\overline{\gamma}^{\mu_2}\ldots\overline{\gamma}^{\mu_n}\overline{\gamma}^{\mu_1})
    = \mathrm{Tr}(\widehat{\gamma}^{\nu_2}\ldots\widehat{\gamma}^{\nu_m}\widehat{\gamma}^{\nu_1}\overline{\gamma}^{\mu_2}\ldots\overline{\gamma}^{\mu_n}\overline{\gamma}^{\mu_1}) \nonumber\\
    &= \mathrm{Tr}\Big(\frac{1}{2}\{\widehat{\gamma}^{\nu_1},\widehat{\gamma}^{\nu_2}\ldots\widehat{\gamma}^{\nu_m}\}\frac{1}{2}\{\overline{\gamma}^{\mu_1},\overline{\gamma}^{\mu_2}\ldots\overline{\gamma}^{\mu_n}\}\Big) \nonumber\\
    &= \sum_{k=2}^{m} \sum_{l=2}^{n} (-1)^k (-1)^l \, \widehat{\eta}^{\nu_1\nu_k} \, \overline{\eta}^{\mu_1\mu_l} \, \mathrm{Tr}(\widehat{\gamma}^{\nu_2}\ldots\widehat{\gamma}^{\nu_{l-1}}\widehat{\gamma}^{\nu_{l+1}}\ldots\widehat{\gamma}^{\nu_{m}}\overline{\gamma}^{\mu_2}\ldots\overline{\gamma}^{\mu_{k-1}}\overline{\gamma}^{\mu_{k+1}}\ldots\overline{\gamma}^{\mu_{n}}) \nonumber\\
    &= \sum_{k=1}^{(m-1)!!} (-1)^{\sigma_k(1,\ldots,m)} \, \widehat{\eta}^{\nu_{\sigma_k(1)}\nu_{\sigma_k(2)}}\ldots\widehat{\eta}^{\nu_{\sigma_k(m-1)}\nu_{\sigma_k(m)}} \\
    &\qquad \times \sum_{l=1}^{(n-1)!!} (-1)^{\sigma_l(1,\ldots,n)} \, \overline{\eta}^{\mu_{\sigma_l(1)}\mu_{\sigma_l(2)}}\ldots\overline{\eta}^{\mu_{\sigma_l(n-1)}\mu_{\sigma_l(n)}} \, \mathrm{Tr}(\mathbb{1}) \nonumber\\
    &= 4 \sum_{k=1}^{(m-1)!!} (-1)^{\sigma_k(1,\ldots,m)} \, \widehat{\eta}^{\nu_{\sigma_k(1)}\nu_{\sigma_k(2)}}\ldots\widehat{\eta}^{\nu_{\sigma_k(m-1)}\nu_{\sigma_k(m)}} \nonumber\\
    &\qquad \times \sum_{l=1}^{(n-1)!!} (-1)^{\sigma_l(1,\ldots,n)} \, \overline{\eta}^{\mu_{\sigma_l(1)}\mu_{\sigma_l(2)}}\ldots\overline{\eta}^{\mu_{\sigma_l(n-1)}\mu_{\sigma_l(n)}}, \nonumber
\end{align}
where we used cyclicity of the trace and $\{\overline{\gamma}^{\mu},\widehat{\gamma}^{\nu}\}=0$,
as well as the fact that only traces with even $n$ and $m$ give rise
to non-vanishing contributions in the second line.
Rewriting the resulting expression using the anticommutator leads to
the third line, while the evaluation of this anticommutator results in
the fourth line.
We again arrive at the penultimate step via recursion 
and obtain the last equality by using \eqref{Eq:GammaTraceDefinitions}.
Comparing Eqs.\ \eqref{Eq:GammaTrace-4-Dim} and \eqref{Eq:GammaTrace-Evanescent}
with Eq.\ \eqref{Eq:GammaTrace-General} leads to the 
desired result given in Eq.\ \eqref{Eq:TraceFactorisationTheorem}
for $\mathbb{\Lambda}=\mathbb{1}$.
Evidently, the additional factor of $1/4$ on the RHS of 
Eq.\ \eqref{Eq:TraceFactorisationTheorem} comes from the
fact that ultimately there are two traces of the identity,
whereas there is only one on the LHS, which needs to be 
taken into account.

This result can be generalised to
$\mathbb{\Lambda}\in\{\mathbb{1},\gamma_5,\projL,\projR\}$,
using the (anti-)commutation relations
of $\gamma_5$ given in Eq.\ \eqref{Eq:BMHV-Algebra-2}
in the first two steps of the treatment of the LHS
of Eq.\ \eqref{Eq:TraceFactorisationTheorem} for 
general $\mathbb{\Lambda}$ (cf.\ Eq.\
\eqref{Eq:GammaTrace-General} for $\mathbb{\Lambda}=\mathbb{1}$), 
as well as
\begin{equation}
    \begin{aligned}
        \mathrm{Tr}(\gamma_5) &= 0,\\
        \mathrm{Tr}(\overline{\gamma}^{\mu}\overline{\gamma}^{\nu}\gamma_5) &= 0,\\
        \mathrm{Tr}(\overline{\gamma}^{\mu_1}\ldots\overline{\gamma}^{\mu_{2n+1}}\gamma_5) &= 0,
    \end{aligned}
\end{equation}
and, for $n=4$,
\begin{equation}
    \begin{aligned}
        \mathrm{Tr}(\gamma_5\overline{\gamma}^{\mu}\overline{\gamma}^{\nu}\overline{\gamma}^{\rho}\overline{\gamma}^{\sigma}) = -4i\varepsilon^{\mu\nu\rho\sigma},
    \end{aligned}
\end{equation}
while even $n\geq6$ can be treated recursively using
\begin{equation}
    \begin{aligned}
        \gamma_5\overline{\gamma}^{\mu}\overline{\gamma}^{\nu}\overline{\gamma}^{\rho} 
        = \overline{\eta}^{\mu\nu}\gamma_5\overline{\gamma}^{\rho}
        - \overline{\eta}^{\mu\rho}\gamma_5\overline{\gamma}^{\nu}
        + \overline{\eta}^{\nu\rho}\gamma_5\overline{\gamma}^{\mu}
        - i\varepsilon^{\mu\nu\rho\sigma}\overline{\gamma}_{\sigma}.
    \end{aligned}
\end{equation}

\section{Explicit Results for the Four-Loop Coefficients}\label{App:ExplicitResults}

In this section, we present the explicit results for all 4-loop 
coefficients used in Sec.\ \ref{Sec:4-Loop-Renormalisation}.
Additionally, we provide the ancillary file \texttt{BMHVAbelian4LoopResults.m},
which contains the full set of Green functions shown in Eqs.\ \eqref{Eq:GF-BB}
to \eqref{Eq:GF-BBBB} and \eqref{Eq:GF-cB} to \eqref{Eq:GF-cBBBB},
along with all coefficients listed below.
We first present the counterterm coefficients for pure gauge boson contributions, 
and then proceed to the fermionic contributions.

\paragraph{Gauge Boson Four-Loop Coefficients:}

We begin with the coefficients associated with the divergent, BRST-invariant bilinear gauge boson counterterm, see Eq.\ \eqref{Eq:CountertermCoeff-GaugeBoson-Inv},
\begin{align}
    \begin{split}
    \mathcal{A}_{BB}^{4,\text{inv}} &= 
    \bigg( \frac{1322627}{97200} + \frac{4019 \, \zeta(3)}{675} \bigg) \text{Tr}\big(\mathcal{Y}_R^8\big)\\
    &\quad + \bigg( \frac{104077}{116640} - \frac{1058 \, \zeta(3)}{2025} \bigg) \text{Tr}\big(\mathcal{Y}_R^6\big) \text{Tr}\big(\mathcal{Y}_R^2\big)\\
    &\quad + \bigg( \frac{537133}{583200} + \frac{34 \, \zeta(3)}{3} \bigg) \text{Tr}\big(\mathcal{Y}_R^4\big)^2
    + \frac{4111}{145800} \, \text{Tr}\big(\mathcal{Y}_R^4\big) \text{Tr}\big(\mathcal{Y}_R^2\big)^2,
    \end{split}\\
    \begin{split}
    \mathcal{B}_{BB}^{4,\text{inv}} &= 
    - \frac{13}{45} \, \text{Tr}\big(\mathcal{Y}_R^8\big)
    - \frac{371}{1080} \, \text{Tr}\big(\mathcal{Y}_R^6\big) \text{Tr}\big(\mathcal{Y}_R^2\big) 
    - \frac{1607}{3240} \, \text{Tr}\big(\mathcal{Y}_R^4\big)^2\\
    &\quad - \frac{707}{14580} \, \text{Tr}\big(\mathcal{Y}_R^4\big) \text{Tr}\big(\mathcal{Y}_R^2\big)^2,
    \end{split}\\
    \begin{split}
    \mathcal{C}_{BB}^{4,\text{inv}} &= 
    \frac{1}{54} \, \text{Tr}\big(\mathcal{Y}_R^6\big) \text{Tr}\big(\mathcal{Y}_R^2\big)
    + \frac{1}{162} \, \text{Tr}\big(\mathcal{Y}_R^4\big)^2
    - \frac{23}{486} \, \text{Tr}\big(\mathcal{Y}_R^4\big) \text{Tr}\big(\mathcal{Y}_R^2\big)^2.
    \end{split}
\end{align}
Next, we present the coefficients for the divergent, BRST-breaking and evanescent counterterm
that is bilinear in gauge bosons, see Eq.\ \eqref{Eq:CountertermCoeff-GaugeBoson-Break-Evanescent},
\begin{align}
    \begin{split}
        \widehat{\mathcal{A}}_{BB}^{4,\text{break}} &=  
        - \bigg( \frac{389961509}{15552000} + \frac{67 \, \pi^4}{9000} + \frac{1119401 \, \zeta(3)}{27000} - \frac{4079 \, \zeta(5)}{45} \bigg) \text{Tr}\big(\mathcal{Y}_R^8\big)\\
        &\quad - \bigg( \frac{19047839}{11664000} + \frac{659 \, \pi^4}{40500} - \frac{8471 \, \zeta(3)}{6750} \bigg) \text{Tr}\big(\mathcal{Y}_R^6\big) \text{Tr}\big(\mathcal{Y}_R^2\big)\\
        &\quad - \bigg( \frac{131895011}{11664000} + \frac{27512 \, \zeta(3)}{675} - \frac{2954 \, \zeta(5)}{45} \bigg) \text{Tr}\big(\mathcal{Y}_R^4\big)^2\\
        &\quad - \frac{102163}{1749600} \text{Tr}\big(\mathcal{Y}_R^4\big) \text{Tr}\big(\mathcal{Y}_R^2\big)^2,
    \end{split}\\
    \begin{split}
        \widehat{\mathcal{B}}_{BB}^{4,\text{break}} &=  
        - \bigg( \frac{582931}{259200} + \frac{67 \, \zeta(3)}{150} \bigg) \text{Tr}\big(\mathcal{Y}_R^8\big) 
        + \bigg( \frac{8263}{12960} - \frac{659 \, \zeta(3)}{675} \bigg) \text{Tr}\big(\mathcal{Y}_R^6\big) \text{Tr}\big(\mathcal{Y}_R^2\big)\\
        &\quad + \frac{361}{2700} \text{Tr}\big(\mathcal{Y}_R^4\big)^2 
        + \frac{1399}{19440} \text{Tr}\big(\mathcal{Y}_R^4\big) \text{Tr}\big(\mathcal{Y}_R^2\big)^2,
    \end{split}\\
    \begin{split}
        \widehat{\mathcal{C}}_{BB}^{4,\text{break}} &= \frac{1231}{4320} \text{Tr}\big(\mathcal{Y}_R^8\big) 
        + \frac{83}{540} \text{Tr}\big(\mathcal{Y}_R^6\big) \text{Tr}\big(\mathcal{Y}_R^2\big) 
        + \frac{7}{1080} \text{Tr}\big(\mathcal{Y}_R^4\big)^2 
        + \frac{\text{Tr}\big(\mathcal{Y}_R^4\big) \text{Tr}\big(\mathcal{Y}_R^2\big)^2}{27},
    \end{split}\\
    \begin{split}
        \widehat{\mathcal{D}}_{BB}^{4,\text{break}} &= - \frac{\text{Tr}\big(\mathcal{Y}_R^8\big)}{72},
    \end{split}
\end{align}
followed by the coefficients for the divergent, BRST-breaking and $4$-dimensional counterterm
that is bilinear in gauge bosons, see Eq.\ \eqref{Eq:CountertermCoeff-GaugeBoson-Break-4dim},
\begin{align}
    \begin{split}
        \overline{\mathcal{A}}_{BB}^{4,\text{break}} &= 
        - \bigg( \frac{2353}{16200} - \frac{4 \, \zeta(3)}{225} \bigg) \text{Tr}\big(\mathcal{Y}_R^8\big)
        - \bigg( \frac{22999}{38880} + \frac{34 \, \zeta(3)}{675} \bigg) \text{Tr}\big(\mathcal{Y}_R^6\big) \text{Tr}\big(\mathcal{Y}_R^2\big)\\ 
        &\quad - \frac{2507}{64800} \text{Tr}\big(\mathcal{Y}_R^4\big)^2 
        - \frac{10327}{1166400} \text{Tr}\big(\mathcal{Y}_R^4\big) \text{Tr}\big(\mathcal{Y}_R^2\big)^2,
    \end{split}\\
    \begin{split}
        \overline{\mathcal{B}}_{BB}^{4,\text{break}} &= 
        \frac{\text{Tr}\big(\mathcal{Y}_R^8\big)}{1080}
        + \frac{13}{3240} \text{Tr}\big(\mathcal{Y}_R^6\big) \text{Tr}\big(\mathcal{Y}_R^2\big)
        + \frac{\text{Tr}\big(\mathcal{Y}_R^4\big)^2}{720}
        + \frac{449}{19440} \text{Tr}\big(\mathcal{Y}_R^4\big) \text{Tr}\big(\mathcal{Y}_R^2\big)^2.
    \end{split}
\end{align}
The coefficient of the finite, symmetry-restoring bilinear gauge boson counterterm reads
\begin{align}\label{Eq:Finite-BB-Break}
    \begin{split}
        \mathcal{F}_{BB}^{4,\text{break}} &= 
        \left(\frac{403759}{25920} + \frac{\pi^4}{3375} + \frac{30113\zeta(3)}{4500} - \frac{403\zeta(5)}{45} \right) \text{Tr}\big(\mathcal{Y}_R^8\big)\\
        &+ \left(\frac{4525151}{11664000} - \frac{17\,\pi^4}{20250} - \frac{15569\zeta(3)}{40500} \right) \text{Tr}\big(\mathcal{Y}_R^6\big)\, \text{Tr}\big(\mathcal{Y}_R^2\big)\\
        &+ \left(\frac{38768057}{7776000} + \frac{12061\zeta(3)}{900} - \frac{713\zeta(5)}{45} \right) \text{Tr}\big(\mathcal{Y}_R^4\big)^2\\
        &- \frac{4240349}{69984000} \text{Tr}\big(\mathcal{Y}_R^4\big) \text{Tr}\big(\mathcal{Y}_R^2\big)^2.
    \end{split}
\end{align}
For the quartic gauge boson counterterm, the coefficients for the 
divergent, BRST-breaking and $4$-dimensional contribution, 
see Eq.\ \eqref{Eq:CountertermCoeff-BBBB-Break}, are provided by 
\begin{align}
    \begin{split}
        \overline{\mathcal{A}}_{BBBB}^{4,\text{break}} &= 
        \left( \frac{1742}{2025} + \frac{8\zeta(3)}{225} \right) \text{Tr}\big(\mathcal{Y}_R^{10}\big)
        + \left( \frac{75677}{48600} + \frac{128\zeta(3)}{135} \right) \text{Tr}\big(\mathcal{Y}_R^8\big) \text{Tr}\big(\mathcal{Y}_R^2\big) \\
        &+ \frac{12359}{3240} \text{Tr}\big(\mathcal{Y}_R^6\big) \text{Tr}\big(\mathcal{Y}_R^4\big)
        - \frac{3539}{291600} \text{Tr}\big(\mathcal{Y}_R^6\big) \text{Tr}\big(\mathcal{Y}_R^2\big)^2
        - \frac{998}{1215} \text{Tr}\big(\mathcal{Y}_R^5\big)^2\\
        &- \frac{407}{4860} \text{Tr}\big(\mathcal{Y}_R^4\big)^2  \, \text{Tr}\big(\mathcal{Y}_R^2\big),
    \end{split}\\
    \begin{split}
        \overline{\mathcal{B}}_{BBBB}^{4,\text{break}} = 
        &- \frac{1}{270} \text{Tr}\big(\mathcal{Y}_R^{10}\big)
        - \frac{19}{810} \text{Tr}\big(\mathcal{Y}_R^8\big) \text{Tr}\big(\mathcal{Y}_R^2\big)
        - \frac{1}{108} \text{Tr}\big(\mathcal{Y}_R^6\big)\text{Tr}\big(\mathcal{Y}_R^4\big) \\
        &- \frac{347}{4860} \text{Tr}\big(\mathcal{Y}_R^6\big)\text{Tr}\big(\mathcal{Y}_R^2\big)^2
        - \frac{2}{3} \text{Tr}\big(\mathcal{Y}_R^4\big)^2 \, \text{Tr}\big(\mathcal{Y}_R^2\big),
    \end{split}
\end{align}
while the coefficient of the finite, symmetry-restoring contribution is given by
\begin{align}\label{Eq:Finite-BBBB-Break}
    \begin{split}
        \mathcal{F}_{BBBB}^{4,\text{break}} = 
        &- \left( \frac{1123939}{18000} - \frac{2\,\pi^4}{3375} + \frac{43376\zeta(3)}{1125} - \frac{392\zeta(5)}{9} \right) \text{Tr}\big(\mathcal{Y}_R^{10}\big) \\
        &+ \left( \frac{3544039}{2916000} + \frac{32\,\pi^4}{2025} - \frac{599\zeta(3)}{2025} \right) \text{Tr}\big(\mathcal{Y}_R^8\big) \text{Tr}\big(\mathcal{Y}_R^2\big) \\
        &- \left( \frac{666463}{388800} + \frac{10888\zeta(3)}{675} + \frac{40\zeta(5)}{9} \right) \text{Tr}\big(\mathcal{Y}_R^6\big) \text{Tr}\big(\mathcal{Y}_R^4\big) \\
        &+ \frac{6224207}{17496000} \text{Tr}\big(\mathcal{Y}_R^6\big) \text{Tr}\big(\mathcal{Y}_R^2\big)^2 
        + \frac{93341}{291600} \text{Tr}\big(\mathcal{Y}_R^4\big)^2\, \text{Tr}\big(\mathcal{Y}_R^2\big) \\
        &+ \left( \frac{730379}{72900} + \frac{7922\zeta(3)}{675} - \frac{536\zeta(5)}{15} \right) \text{Tr}\big(\mathcal{Y}_R^5\big)^2.
    \end{split}
\end{align}

\paragraph{Fermion Four-Loop Coefficients:}

Analogous to the case of the gauge boson contributions, we proceed to list all
coefficients for the fermionic 4-loop counterterms.
We begin with those for the divergent fermionic contributions, 
see Eq.\ \eqref{Eq:CountertermCoeff-Fermion-Inv},
\begin{align}
    \begin{split}
        \mathcal{A}_{\psi\overline{\psi}, ij}^{4,\text{inv}} &= 
        \bigg( \frac{156477439}{2592000} + \frac{32773 \zeta(3)}{300} - 160 \zeta(5) \bigg) \big( \mathcal{Y}_R^8 \big)_{ij}\\
        &\quad - \bigg( \frac{14415971}{2592000} + \frac{2357 \zeta(3)}{450} \bigg) \text{Tr} \big( \mathcal{Y}_R^2 \big) \big( \mathcal{Y}_R^6 \big)_{ij}\\
        &\quad -  \bigg[  
        \bigg( \frac{28122617}{1555200} - \frac{3104 \zeta(3)}{225} \bigg) \text{Tr} \big( \mathcal{Y}_R^4 \big)  
        - \frac{3213671}{11664000} \text{Tr} \big( \mathcal{Y}_R^2 \big)^2 \bigg] \big( \mathcal{Y}_R^4 \big)_{ij}\\
        &\quad -
        \bigg( \frac{8177}{2880} - \frac{264 \zeta(3)}{25} \bigg) \text{Tr} \big( \mathcal{Y}_R^5 \big) \big( \mathcal{Y}_R^3 \big)_{ij}\\
        &\quad - 
        \bigg[  
        \bigg( \frac{2252129}{777600} - \frac{547 \zeta(3)}{450} \bigg) \text{Tr} \big( \mathcal{Y}_R^6 \big)
        + \frac{66841}{11664000} \text{Tr} \big( \mathcal{Y}_R^2 \big)^3\\
        &\qquad + \bigg( \frac{161755}{186624} + \frac{1069 \zeta(3)}{450} \bigg) \text{Tr} \big( \mathcal{Y}_R^4 \big) \text{Tr} \big( \mathcal{Y}_R^2 \big) 
        \bigg] \big( \mathcal{Y}_R^2 \big)_{ij} \\
        &\quad + 
        \bigg[
        \bigg( \frac{49367}{9600} - \frac{2 \zeta(3)}{5} \bigg) \text{Tr} \big( \mathcal{Y}_R^7 \big)\\
        &\qquad + \bigg( \frac{1271}{1440} - \frac{257 \zeta(3)}{75} \bigg) \text{Tr} \big( \mathcal{Y}_R^5 \big)  \text{Tr} \big( \mathcal{Y}_R^2 \big) \bigg] \big( \mathcal{Y}_R \big)_{ij} ,
    \end{split}\\
    \begin{split}
        \mathcal{B}_{\psi\overline{\psi}, ij}^{4,\text{inv}} &= 
        \frac{29063}{4800} \big( \mathcal{Y}_R^8 \big)_{ij}
        - \frac{147353}{43200} \text{Tr} \big( \mathcal{Y}_R^2 \big) \big( \mathcal{Y}_R^6 \big)_{ij}\\
        &\quad - \bigg( \frac{1}{90} \text{Tr}\big( \mathcal{Y}_R^4 \big) - \frac{43811}{64800} \text{Tr}\big( \mathcal{Y}_R^2 \big)^2 \bigg) \big( \mathcal{Y}_R^4 \big)_{ij} 
        + \frac{299}{240} \text{Tr}\big( \mathcal{Y}_R^5 \big)  \big( \mathcal{Y}_R^3 \big)_{ij}\\
        &\quad - \bigg( \frac{611}{12960} \text{Tr}\big( \mathcal{Y}_R^6 \big) 
        - \frac{13429}{38880} \text{Tr} \big( \mathcal{Y}_R^4 \big)  \text{Tr} \big( \mathcal{Y}_R^2 \big) 
        - \frac{157}{194400}  \text{Tr}\big( \mathcal{Y}_R^2 \big)^3 \bigg) \big( \mathcal{Y}_R^2 \big)_{ij} \\
        &\quad  - \bigg( \frac{37}{480} \text{Tr}\big( \mathcal{Y}_R^7 \big) 
        + \frac{191}{360} \text{Tr} \big( \mathcal{Y}_R^5 \big) \text{Tr} \big( \mathcal{Y}_R^2 \big) \bigg) \big( \mathcal{Y}_R \big)_{ij},
    \end{split}\\
    \begin{split}
        \mathcal{C}_{\psi\overline{\psi},ij}^{4,\text{inv}} &= 
        \frac{451}{720} \big( \mathcal{Y}_R^8 \big)_{ij}
        - \frac{1057}{2160} \text{Tr}\big( \mathcal{Y}_R^2 \big)  \big( \mathcal{Y}_R^6 \big)_{ij}\\
        &\quad - \bigg( \frac{7}{432} \text{Tr}\big( \mathcal{Y}_R^4 \big)
        - \frac{109}{540} \text{Tr}\big( \mathcal{Y}_R^2 \big)^2 \bigg) \big( \mathcal{Y}_R^4 \big)_{ij}\\
        &\quad + \bigg( \frac{\text{Tr}\big( \mathcal{Y}_R^6 \big)}{216}
        + \frac{\text{Tr}\big( \mathcal{Y}_R^4 \big) \text{Tr}\big( \mathcal{Y}_R^2 \big)}{432}
        + \frac{\text{Tr}\big( \mathcal{Y}_R^2 \big)^3}{3240} \bigg) \big( \mathcal{Y}_R^2 \big)_{ij},
    \end{split}\\
    \begin{split}
        \mathcal{D}_{\psi\overline{\psi},ij}^{4,\text{inv}} &= \frac{1}{24} \big( \mathcal{Y}_R^8 \big)_{ij},
    \end{split}
\end{align}
as well as those in Eq.\ \eqref{Eq:CountertermCoeff-Fermion-Break},
\begin{align}
    \begin{split}
        \overline{\mathcal{A}}_{\psi\overline{\psi}, ij}^{4,\text{break}} &= 
        - \bigg( \frac{29481469}{1296000} + \frac{5461}{900} \zeta(3) \bigg)  
        \big( \mathcal{Y}_R^8 \big)_{ij}\\  
        &+  
        \bigg( \frac{82599047}{7776000} + \frac{2411}{900} \zeta(3) \bigg) 
        \text{Tr} \big( \mathcal{Y}_R^2 \big) \big( \mathcal{Y}_R^6 \big)_{ij}\\  
        &-  
        \bigg[ 
        \bigg( \frac{919571}{388800} + \frac{1198}{225} \zeta(3) \bigg) \text{Tr} \big( \mathcal{Y}_R^4 \big) 
        + \frac{10432717}{11664000} \text{Tr} \big( \mathcal{Y}_R^2 \big)^2 
        \bigg]
        \big( \mathcal{Y}_R^4 \big)_{ij}\\  
        &+  
        \bigg( \frac{8233}{14400} - \frac{707}{100} \zeta(3) \bigg)
        \text{Tr} \big( \mathcal{Y}_R^5 \big) 
        \big( \mathcal{Y}_R^3 \big)_{ij}\\  
        &+ \bigg[  
        \bigg( \frac{1194907}{777600} - \frac{23}{90} \zeta(3) \bigg)
        \text{Tr} \big( \mathcal{Y}_R^6 \big)
        - \frac{22139}{1296000} \text{Tr} \big( \mathcal{Y}_R^2 \big)^3\\  
        &\quad +   
        \bigg( \frac{41257}{186624} + \frac{335}{108} \zeta(3) \bigg)
        \text{Tr} \big( \mathcal{Y}_R^4 \big) \text{Tr} \big( \mathcal{Y}_R^2 \big)
        \bigg] \big( \mathcal{Y}_R^2 \big)_{ij}\\  
        &- \bigg[  
        \bigg( \frac{49367}{9600} - \frac{2}{5} \zeta(3) \bigg)
        \text{Tr} \big( \mathcal{Y}_R^7 \big)\\
        &\quad + \bigg( \frac{1271}{1440} - \frac{257}{75} \zeta(3) \bigg)
        \text{Tr} \big( \mathcal{Y}_R^5 \big) \text{Tr} \big( \mathcal{Y}_R^2 \big)
        \bigg] \big( \mathcal{Y}_R \big)_{ij},
    \end{split}\\
    \begin{split}
        \overline{\mathcal{B}}_{\psi\overline{\psi}, ij}^{4,\text{break}} &= 
        - \frac{16609}{7200} \big( \mathcal{Y}_R^8 \big)_{ij}
        + \frac{254021}{129600} \text{Tr} \big( \mathcal{Y}_R^2 \big) \big( \mathcal{Y}_R^6 \big)_{ij}
        - \frac{149}{240} \text{Tr} \big( \mathcal{Y}_R^5 \big) \big( \mathcal{Y}_R^3 \big)_{ij}\\
        &\quad +  
        \bigg( \frac{121}{810} \text{Tr} \big( \mathcal{Y}_R^4 \big) - \frac{145091}{194400} \text{Tr} \big( \mathcal{Y}_R^2 \big)^2 \bigg) \big( \mathcal{Y}_R^4 \big)_{ij}\\
        &\quad -  
        \bigg( \frac{719}{12960} \text{Tr} \big( \mathcal{Y}_R^6 \big)
        + \frac{107}{864} \text{Tr} \big( \mathcal{Y}_R^4 \big) \text{Tr} \big( \mathcal{Y}_R^2 \big)
        - \frac{2909}{64800} \text{Tr} \big( \mathcal{Y}_R^2 \big)^3 \bigg) 
        \big( \mathcal{Y}_R^2 \big)_{ij}\\
        &\quad + 
        \bigg( \frac{37}{480} \text{Tr} \big( \mathcal{Y}_R^7 \big)  
        + \frac{191}{360} \text{Tr} \big( \mathcal{Y}_R^5 \big) \text{Tr} \big( \mathcal{Y}_R^2 \big) \bigg) 
        \big( \mathcal{Y}_R \big)_{ij},   
    \end{split}\\
    \begin{split}
        \overline{\mathcal{C}}_{\psi\overline{\psi}, ij}^{4,\text{break}} &= 
        - \frac{31}{360} \big( \mathcal{Y}_R^8 \big)_{ij} 
        + \frac{41}{720} \text{Tr} \big( \mathcal{Y}_R^2 \big) \big( \mathcal{Y}_R^6 \big)_{ij}\\
        &\quad - \bigg( \frac{\text{Tr} \big( \mathcal{Y}_R^4 \big)}{54}  
        + \frac{173}{3240} \text{Tr} \big( \mathcal{Y}_R^2 \big)^2 \bigg) \big( \mathcal{Y}_R^4 \big)_{ij}\\
        &\quad + 
        \bigg( \frac{1}{216} \text{Tr} \big( \mathcal{Y}_R^6 \big)  
        + \frac{5}{432} \text{Tr} \big( \mathcal{Y}_R^4 \big) \text{Tr} \big( \mathcal{Y}_R^2 \big)
        + \frac{37}{1080} \text{Tr} \big( \mathcal{Y}_R^2 \big)^3 \bigg) \big( \mathcal{Y}_R^2 \big)_{ij}. 
    \end{split}
\end{align}
Finally, the coefficient of the finite, symmetry-restoring fermionic counterterm
is given by
\begin{align}\label{Eq:Finite-FF-Break}
    \begin{split}
        \mathcal{F}_{\overline{\psi}\psi, ij}^{4,\text{break}} &=
        - \bigg( 
        \frac{2296950361}{25920000} - \frac{113 \pi^4}{18000} + \frac{3803231}{54000} \zeta(3) - \frac{\zeta(5)}{15} 
        \bigg) \big( \mathcal{Y}_R^8 \big)_{ij} \\  
        &+ \bigg( 
        \frac{887862323}{155520000} + \frac{2411 \pi^4}{54000} + \frac{113087}{18000} \zeta(3) 
        \bigg) \text{Tr} \big( \mathcal{Y}_R^2 \big) \big( \mathcal{Y}_R^6 \big)_{ij}\\  
        &+ \bigg[   
        \bigg( \frac{33905827}{583200} - \frac{37 \pi^4}{3375} + \frac{307901}{6750} \zeta(3) - \frac{7153}{45} \zeta(5) \bigg) 
        \text{Tr} \big( \mathcal{Y}_R^4 \big)\\
        &\quad + \bigg( \frac{579755521}{699840000} + \frac{29}{540} \zeta(3) \bigg)
        \text{Tr} \big( \mathcal{Y}_R^2 \big)^2 
        \bigg] \big( \mathcal{Y}_R^4 \big)_{ij} \\  
        &+ \bigg( 
        \frac{1593257}{172800} - \frac{7 \pi^4}{6000} - \frac{729797}{6000} \zeta(3) + \frac{479}{6} \zeta(5) 
        \bigg) 
        \text{Tr} \big( \mathcal{Y}_R^5 \big) \big( \mathcal{Y}_R^3 \big)_{ij}\\  
        &+ \bigg[ 
        \bigg( \frac{975213329}{46656000} - \frac{1433 \pi^4}{81000} + \frac{811729}{27000} \zeta(3) - \frac{1141}{15} \zeta(5) \bigg) \text{Tr} \big( \mathcal{Y}_R^6 \big) \\
        &\quad +
        \bigg( \frac{78192881}{139968000} + \frac{67 \pi^4}{1296} + \frac{299}{400} \zeta(3) \bigg) 
        \text{Tr} \big( \mathcal{Y}_R^4 \big) \text{Tr} \big( \mathcal{Y}_R^2 \big)\\  
        &\quad -  
        \bigg( \frac{3350431}{25920000} - \frac{37}{540} \zeta(3) \bigg)
        \text{Tr} \big( \mathcal{Y}_R^2 \big)^3
        \bigg] \big( \mathcal{Y}_R^2 \big)_{ij}\\  
        &+ \bigg[  
        \bigg( \frac{42763633}{1728000} + \frac{\pi^4}{150} - \frac{5923}{300} \zeta(3) - \frac{263}{10} \zeta(5) \bigg)
        \text{Tr} \big( \mathcal{Y}_R^7 \big)\\  
        &\quad - 
        \bigg( \frac{3833903}{1296000} - \frac{257 \pi^4}{4500} + \frac{1261}{1500} \zeta(3) \bigg) 
        \text{Tr} \big( \mathcal{Y}_R^5 \big) \text{Tr} \big( \mathcal{Y}_R^2 \big)
        \bigg] \big( \mathcal{Y}_R \big)_{ij}.
    \end{split}
\end{align}

\end{appendices}
% -------------------------------------------------------------------------

%%%%% Bibliography %%%%%
\printbibliography

\end{document}